\documentclass[aps,reprint]{revtex4-2}
\usepackage{etoolbox} %
\usepackage{blindtext}
\usepackage{graphicx}
\usepackage{subcaption}
\usepackage{amssymb}
\usepackage{xcolor}
\usepackage{graphicx}
\usepackage{floatrow}
\usepackage[fleqn]{amsmath}
\usepackage[font=small,skip=3pt]{caption}
\usepackage{makecell}
\usepackage{hyperref} 
\usepackage{cleveref}
\usepackage{url}
\usepackage{siunitx}
\makeatletter
\appto{\appendix}{%
  \@ifstar{\def\theequation@prefix{A.}}%
          {}%
}
\makeatother

\begin{document}
\title{Revisiting Fusion in D-${}^{3}$He Plasmas With Spin-Polarized Fuel}
\author{J. F. Parisi$^{1}$}
\email{jparisi@pppl.gov}
\author{A. Diallo$^1$}
\author{S. Meschini$^2$}
\affiliation{$^1$Princeton Plasma Physics Laboratory, Princeton University, Princeton, NJ, USA}
\affiliation{$^2$Polytechnic University of Turin, Turin, Piedmont, Italy}
\begin{abstract}
Spin-polarized fuel (SPF) is recognized for enhancing fusion reactivity, but it could provide other advantages particularly relevant to advanced fusion fuels. In this work, we calculate how SPF in D-${}^3$He plasmas affects not only D-${}^3$He fusion reactions but D-D fusion and subsequent secondary reactions. By incorporating multiple effects, we show how, under optimistic assumptions, the fusion power relative to unpolarized D-${}^3$He fusion power could increase by more than a factor of three by polarizing the deuterium and helium-3. Such an increase may improve the feasibility of fusion concepts using D-${}^3$He fuel. We perform a case study with a hypothetical pulsed magneto-inertial fusion device using polarized D-${}^3$He fuel, showing how the net electric power could increase by almost an order of magnitude. We also consider the potential of SPF to provide a path to fully aneutronic fusion. This work is a new look at how SPF could improve the feasibility of fusion concepts using advanced fuels.
\end{abstract}
\maketitle

\section{Introduction}

An attractive fuel for fusion power plants is deuterium-tritium (D-T) due to its high fusion reactivity at relatively low temperature ($T \sim$ 10-20 keV) \cite{Strachan1994short,Keilhacker1999,Wurzel2022}. At this temperature, the D-T fusion reactivity is at least two orders of magnitude higher than for any other known fusion reactions such as deuterium-deuterium (D-D) and deuterium-helium-3 (D-${}^{3}$He). Therefore, in a thermonuclear D-T plasma at $T \sim$ 10-20 keV, typically more than 99\% of the fusion power comes from D-T reactions \cite{Strachan1994short,Keilhacker1999,Bonheure_2012}. Because of the pessimistic scaling of radiative and transport losses with higher temperature \cite{Rider_1995,Rider_1997}, most magnetic confinement power plant designs plan to use D-T fuel at relatively low temperature in the plasma core ($T \sim$ 10-20 keV) \cite{Lawson1957,Mau1999,Sethian_2005,Najmabadi_2006,Najmabadi2008,Meier2014,Sorbom2015,Menard2016,Federici2019b,Buttery2021,Wade2021,Fradera2021,Schoofs2022,Morris_2022,Menard2023_IAEA,Muldrew2024}.

In a higher temperature plasma ($T \gtrsim$ 100 keV), the D-T, D-D, and D-${}^{3}$He reactivities are within an order of magnitude of each other, and at $T \sim$ 1000 keV the reactivities are almost equal. However, producing net electricity at these higher temperatures is extremely challenging using any fusion fuel because of prohibitively high bremsstrahlung losses \cite{Rider_1995,Rider_1997}.

If net electric power were somehow able to be generated from advanced fuels such D-D, and D-${}^{3}$He, there would be some significant advantages over D-T. First, dealing with 14 MeV D-T neutrons in a fusion power plant is challenging \cite{Gilbert_2011,Fischer_2015}. The D-T neutrons break down and activate power plant materials, requiring hundreds of years of safe storage after the plant is decommissioned. There are also more engineering solutions for handling 2.5 MeV neutrons from the D+D $\to$ n+${}^3$He reaction than 14 MeV neutrons from D-T fusion. Furthermore, if a pure aneutronic burn can be somehow achieved, neutron shielding can be removed, significantly simplifying the power plant design. Second, tritium is radioactive with a half-life of 12 years and difficult to produce \cite{Kaufman1954,Kovari2018}. Therefore, the absence of 14 MeV D-T fusion neutrons and a much lower - or entirely absent - tritium inventory reduces engineering and regulation complexity \cite{Roth2008}. Third, the possibility of direct energy conversion (DEC) \cite{rosa1987magnetohydrodynamic,tomita1993direct,rostoker1997colliding,wurden2016magneto} with much higher electricity conversion efficiency ($\sim$80+\%) than standard thermal cycles ($\sim$30-40\%) can significantly increase the net electric power output, although there are concerns that charged particles may lose most of their energy before direct converting to electricity \cite{Stott_2005}, limiting the practical effectiveness of DEC. Fourth, some advanced fuels such as D-D and p-${}^{11}$B are much easier to supply than the lithium required for D-T \cite{Fasel_2005,Meschini2023,Meschini_2025} (although there are advanced fuels such as D-${}^{3}$He \cite{Wittenberg_1992,Shea2010_helium,Simko2014_lunar,Meschini_2025} with major fuel supply challenges).

\begin{figure*}[tb!]
    \centering
    \begin{subfigure}[t]{0.9\textwidth}
    \centering
    \includegraphics[width=1.0\textwidth]{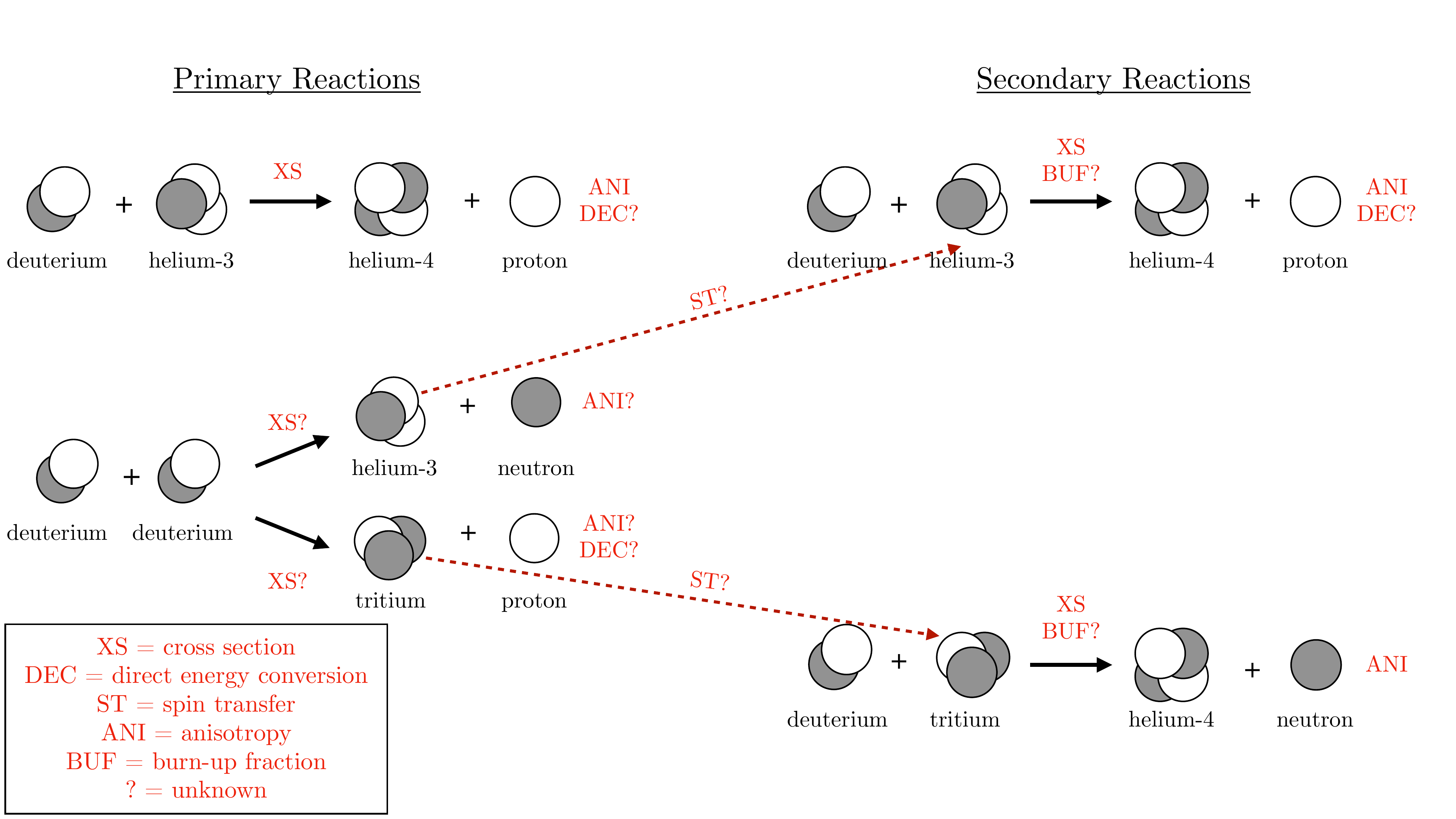}
    \end{subfigure}
    \caption{Diagram for dominant fusion reactions in D-${}^{3}$He plasmas. The effect of spin-polarized fuel on different components are in red, question marks indicate significant uncertainty on the effects of spin-polarized fuel.}
    \label{fig:SPF_DHe3}
\end{figure*}

The drawbacks of D-T fusion have motivated efforts to overcome the barriers in advanced fuels identified in earlier analyses \cite{Rider_1995}. These approaches typically focus on either reducing energy losses or enhancing fusion power output. Loss-reduction strategies include fast proton heating to selectively boost ion temperatures and minimize electron heating and radiation losses \cite{Ochs_2024,Rostoker_2003,Kolmes_2022}, alpha channeling, which extracts energy from fusion-produced alpha particles \cite{Herrmann_1997} before thermalization, thus lowering bremsstrahlung losses and increasing the main ion temperature \cite{Fisch_1992,Ochs_2022}, and alpha particle de-mixing, where fusion ash is separated from reacting fuels to improve energy balance \cite{Ochs_2025}. Fusion power enhancement strategies include optimized fast-ion distributions driven by beam injection and wave-particle interactions \cite{Magee_2019}, non-Maxwellian particle distributions engineered to increase reactivity \cite{Hay_2015,Xie_2023}, advanced ignition schemes \cite{Labaune_2013,Baccou_2015}, and alternative concepts such as magnetized target fusion, field-reversed configurations, and z-pinches \cite{Basko_2000,Cohen_2000,Rostoker_2003,Binderbauer_2010,Ryzhkov_2023}. 

In this work, we re-visit a complementary avenue for improving the feasibility of advanced fusion fuels: using spin-polarized fuel to increase the fusion power in D-${}^{3}$He plasmas \cite{Kulsrud1986,Mitarai1992}. By aligning the spins of the deuterium and helium-3 nuclei, the cross section increases by roughly 50\% and the fusion product particle emission is no longer necessarily isotropic \cite{Kulsrud1986,Spiliotis_2021}. Forthcoming experiments on the DIII-D tokamak will test the polarization lifetime of deuterium and helium-3 \cite{Garcia2023,Baylor2023,Heidbrink2024,Garcia2025}.

The central idea presented in this paper is as follows. While polarizing deuterium and helium-3 is known to increase the D-${}^{3}$He reactivity by $\sim$50\% \cite{Kulsrud1986}, an even larger increase in fusion power might be achieved with the following phenomena: (1) increased D-D reactivity with spin-polarized deuterium, (2) increased burn-up fraction of secondary D-T and D-${}^{3}$He products born at high energy, and (3) subsequent optimization of the deuterium-helium-3 fuel ratio. (1) and (2) are an enhancement of the catalyzed D-D cycle \cite{Albert_1956,Miley_1977}. Further increases in the net electric power might be achieved by a DEC efficiency that also improves with polarization due to anisotropic emission spectra and further deuterium-helium-3 fuel ratio optimization for net electric power (not fusion power) \cite{Parisi_2025b}. There are additional effects of using spin-polarized fuel -- such as increased alpha heating increasing plasma temperature and therefore reactivity \cite{Smith2018IAEA,Heidbrink2024} -- that we do not consider here, but may further modify the fusion power.

The use of SPF adds further complexity when at least two distinct fusion reactions are dynamically important. In contrast to D-T plasmas at $T \sim 10-20$ keV - where secondary reactions are relatively small and D-T fusion reactions dominate - in D-${}^{3}$He plasmas, the effects of SPF are more complex because of additional D-D fusion reactions and because the tritium and helium-3 produced from D-D reactions are fusion fuels themselves. Therefore, the beneficial effects of SPF can cascade down to secondary D-T and D-${}^{3}$He reactions. Further complexity and uncertainty arises because the effects of polarization on D-D reactivity remain poorly characterized, particularly the effect of SPF on the D-D cross section and how spin is transferred from D-D to the fusion products tritium and helium-3 that affects subsequent secondary D-T and D-${}^{3}$He reactions. While the impact of spin polarization on D-T and D-${}^{3}$He reactivities is relatively temperature-independent at fusion-relevant temperatures \cite{Kulsrud1986}, the reactivities of the two D-D branches are predicted to exhibit strong temperature dependence \cite{Lemaitre1993,Engels2003,Deltuva2007,Paetz2010}. Beyond modifying fusion reactivities, SPF can also alter emission directions of fusion products. For the aneutronic D-${}^{3}$He reaction, this may influence the efficiency of converting charged fusion products into electricity via direct energy conversion. Finally, the effect of spin-polarization on the burn-up fraction in secondary D-${}^{3}$He and D-T fusion reactions is also unclear and will depend strongly on the fusion concept. All of these effects are shown schematically in \Cref{fig:SPF_DHe3}. Consequently, for all the reasons listed above, spin-polarized fuel could have a substantial impact on the feasibility and performance of fusion systems using D-${}^{3}$He fuel and direct energy conversion. Predicting the effect of SPF on fusion power plants using D-${}^{3}$He plasmas is far more complicated than for power plants using D-T plasmas.

\begin{figure}[bt!]
    \centering
    \begin{subfigure}[t]{0.88\textwidth}
    \centering
    \includegraphics[width=1.00\textwidth]{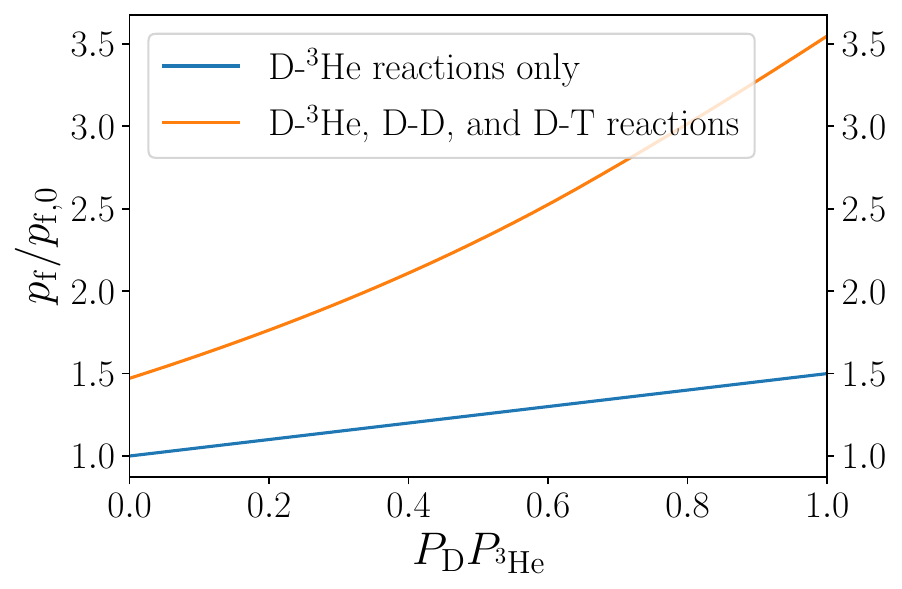}
    \caption{}
    \end{subfigure}
    \centering
    \begin{subfigure}[t]{0.88\textwidth}
    \centering
    \includegraphics[width=1.00\textwidth]{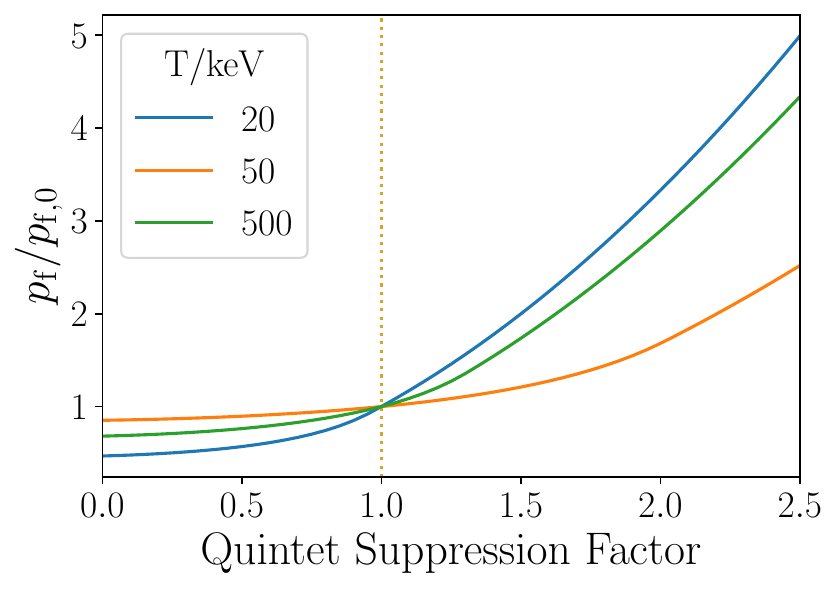}
    \caption{}
    \end{subfigure} 
    \caption{(a) Normalized fusion power $p_\mathrm{f}$ versus deuterium and helium-3 vector polarization product $P_\mathrm{D} P_\mathrm{{}^{3}He}$. (b) Fusion power versus Quintet Suppression Factor. In (a) we assumed a Quintet Suppression Factor = 1.5, plasma temperature T = 50 keV, and the normalizing power $p_\mathrm{f,0}$ corresponds to the power only from unpolarized D-${}^{3}$He reactions in a D-${}^{3}$He plasma. In (b) we assumed a fully polarized D-${}^{3}$He plasma: $P_\mathrm{D} P_\mathrm{{}^{3}He}=1.0$ and the normalizing normalizing power $p_\mathrm{f,0}$ corresponds to a fully-polarized plasma with Quintet Suppression Factor = 1.0.}
    \label{fig:pDelta_summary_QSF}
\end{figure}

For readers seeking a quick summary, we plot two important results in \Cref{fig:pDelta_summary_QSF}, where we compare the thermal fusion power in a D-${}^{3}$He plasma with different levels of polarization in a scenario with highly optimistic assumptions. In \Cref{fig:pDelta_summary_QSF} (a) we plot the normalized fusion power versus the deuterium and helium-3 vector polarization $P_\mathrm{D} P_\mathrm{{}^{3}He}$. When $P_\mathrm{D} P_\mathrm{{}^{3}He} = 1$, all of the deuterium and helium-3 nuclear spins are aligned, and when $P_\mathrm{D} P_\mathrm{{}^{3}He} = 0$, there is zero net alignment. In \Cref{fig:pDelta_summary_QSF} (a), the quantity $p_\mathrm{f}/p_\mathrm{f,0}$ is the fusion power density $p_\mathrm{f}$ relative to the base-case $p_\mathrm{f,0}$ considering only D-${}^{3}$He reactions with zero spin-polarization. Higher values of $P_\mathrm{D} P_\mathrm{{}^{3}He}$ correspond to more spin polarization: $P_\mathrm{D} P_\mathrm{{}^{3}He} = 0$ is unpolarized and $P_\mathrm{D} P_\mathrm{{}^{3}He} = 1$ is fully polarized. Relative to an unpolarized fusion plasma with only D-${}^{3}$He reactions, a fully polarized D-${}^{3}$He plasma with a Quintet Suppression Factor \cite{Paetz2010,Deltuva_2010,Grigoryev_2011,Engels_2014,Viviani_2023}
\begin{equation}
    \mathrm{QSF} \equiv \frac{\sigma_{1,1}}{\overline{\sigma}}
\end{equation}
of 2.5 has 3.6 times the fusion power of unpolarized D-${}^{3}$He fusion reactions when primary D-${}^{3}$He, D-D and secondary D-${}^{3}$He, D-T reactions are included. The QSF is the decrease/increase in the fusion cross section $\sigma$ for D-D reactions when the D-D spins are aligned ($\sigma_{1,1}$) versus when the D-D is unpolarized ($\overline{\sigma}$). There is significant uncertainty in the value of QSF, from QSF$\simeq1/10$ to QSF$\simeq$2.5 \cite{Engels_2014}. This surprisingly large increase in fusion power arises from two effects: (1) the increase in fusion reactivity, (2) increase in D-T and D-${}^{3}$He burnup fraction in secondary reactions, and (3) an optimization in the D-${}^{3}$He fuel ratio. In \Cref{fig:pDelta_summary_QSF} (b), we plot the dependence of $p_\mathrm{f}/p_\mathrm{f,0}$ on the QSF for a fully polarized D-${}^{3}$He plasma for three temperatures with $P_\mathrm{D} P_\mathrm{{}^{3}He}=1.0$ where $p_\mathrm{f,0}$ corresponds to the QSF=1 case. In the optimistic case that SPF increases the D-D cross section by 150\% (QSF=2.5), the fusion power increases by $\sim$100-400\% relative to the nominal QSF =1 case, depending on the plasma temperature. The fusion power is more sensitive to the QSF at temperatures where most of the fusion power comes from D-D reactions.

The general result is that the power density of a spin-polarized D-${}^{3}$He plasma and all of its secondary reactions can be comparable to the power density of D-T reactions from a D-T plasma at temperatures $T \gtrsim 100$ keV. Once the effects of direct energy conversion, radiation, and transport are accounted for, under optimistic assumptions of spin-polarized fuel, a fusion power plant using D-${}^{3}$He fuel may generate significant net electric power. Thus, using spin-polarized fuel could increase the feasibility of D-${}^{3}$He power plants. Including radiative and transport effects is outside the scope of this work, but we plan to address these important questions in future work.

The layout of this work is as follows. In \Cref{sec:SPF_intro} we introduce spin-polarized fuel. The paper is then split into two general areas. The first area in \Cref{sec:steadystatemodel} addresses the various effects of SPF in a steady-state fusion model of a D-${}^{3}$He plasma. The second area in \Cref{sec:reactivity_energyrecovery} addresses the combined effects on engineering gain of increased fusion reactivity and increasing energy recovery using SPF. We summarize in \Cref{sec:discussion}. 

Additional details of the fusion reactions studied in this paper and the effect of including T-${}^3$He and ${}^3$He-${}^3$He reactions are in \Cref{app:reactions}.

\section{Spin-Polarized Fuel} \label{sec:SPF_intro}

\begin{figure}[tb!]
    \centering
    \begin{subfigure}[t]{0.97\textwidth}
    \centering
    \includegraphics[width=1.0\textwidth]{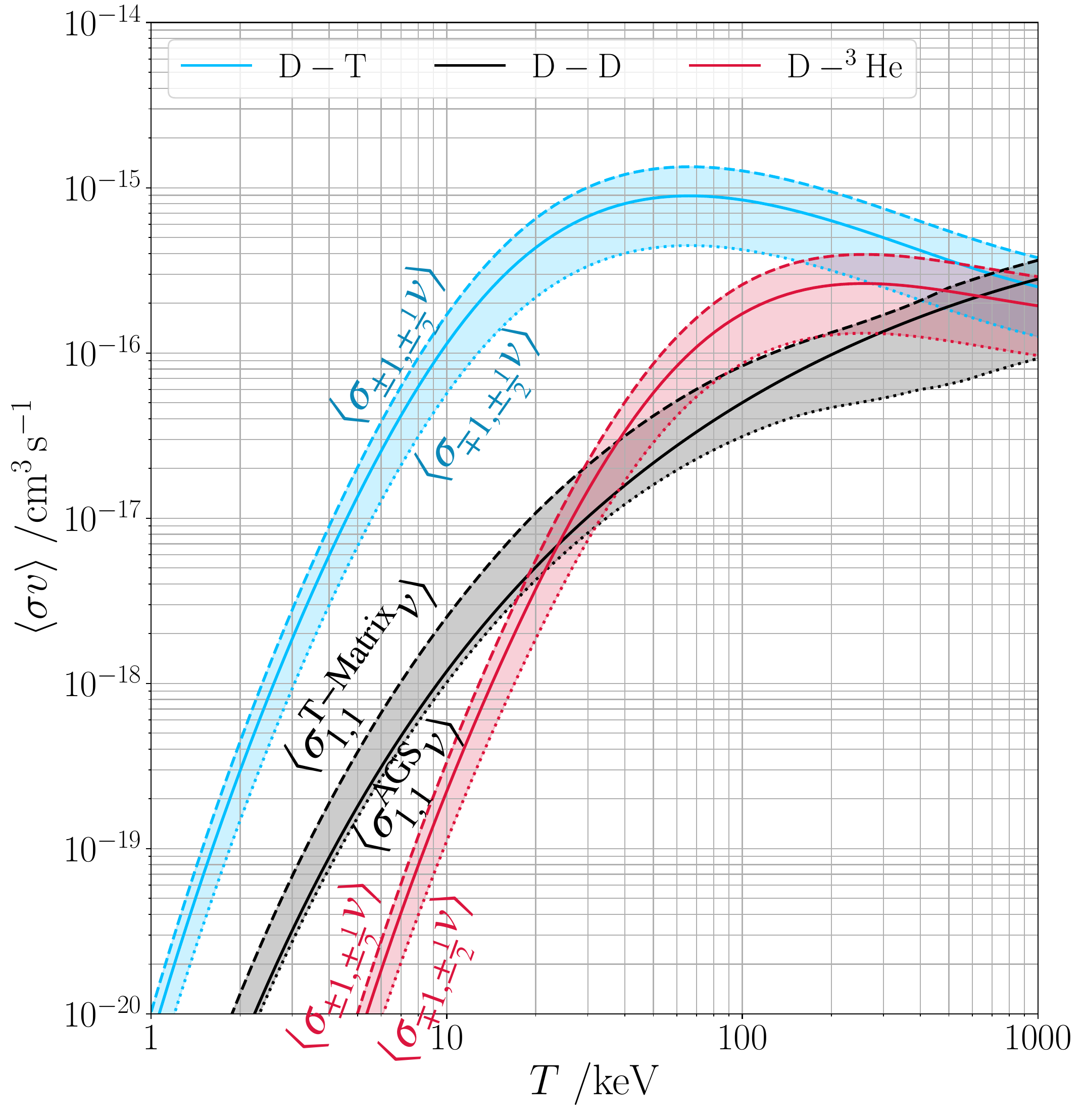}
    \end{subfigure}
    \caption{Reactivities for D-T, D-${}^{3}$He, and D-D with unpolarized (solid lines) and polarized (dashed and dotted) fuel. See text for more details.}
    \label{fig:reactivities_polarized}
\end{figure}

In this section, we introduce our model for the effect of spin-polarized fuel on fusion reactivity. We mainly consider the following fusion reactions,
\begin{align}
\text{(1) } & \mathrm{D} + {}^3\mathrm{He} \longrightarrow \alpha + \mathrm{p} \;\; \left( E_{\mathrm{D{}^{3}He}} = 18.4 \; \mathrm{MeV} \right) \\
\text{(2a) } & \mathrm{D} + \mathrm{D} \longrightarrow {}^3\mathrm{He} + \mathrm{n}
\; \left( E_{\mathrm{DD,n}} = 3.3 \; \mathrm{MeV} \right) \\
\text{(2b) } & \mathrm{D} + \mathrm{D} \longrightarrow \mathrm{T} + \mathrm{p}
\;\;\;\; \left( E_{\mathrm{DD,p}} = 4.0 \; \mathrm{MeV} \right) \\
\text{(3) } & \mathrm{D} + \mathrm{T} \longrightarrow \alpha + \mathrm{n} \;\;\;\;\; \left( E_{\mathrm{DT}} = 17.6 \; \mathrm{MeV} \right)
\end{align}
where $E_{\mathrm{D{}^{3}He}}$, $E_{\mathrm{DD,n}}$, $E_{\mathrm{DD,p}}$, and $E_{\mathrm{DT}}$ are the energies released from the fusion reactions.

For D-${}^{3}$He fusion, the ratio of the polarized cross section $\sigma_{P_\mathrm{D} P_\mathrm{{}^{3}He}}$ to a nominal unpolarized cross section $\overline{\sigma}$ is approximately \cite{Kulsrud1986}
\begin{equation}
\frac{\sigma_{P_\mathrm{D} P_\mathrm{{}^{3}He}}}{\overline{\sigma}} = 1 + \frac{P_\mathrm{D} P_\mathrm{{}^{3}He}}{2}.
\label{eq:AJ2}
\end{equation}
For unpolarized fusion, the nuclear spins are randomly oriented giving $P_\mathrm{D}=P_\mathrm{{}^{3}He}=0$. Here, $P_\mathrm{D}, P_\mathrm{{}^{3}He}$ are the vector polarizations of deuterium and helium-3, where $P_\mathrm{D} = D_{1} - D_{-1}$ and $P_\mathrm{{}^{3}He} = \mathrm{{}^{3}He}_{1/2} - \mathrm{{}^{3}He}_{-1/2}$, and, $D_m$ and $\mathrm{{}^{3}He}_m$ are the probabilities of being in a nuclear spin state $m$, where $m = 1,0,-1$ for deuterium and $m = 1/2,-1/2$ for helium-3, satisfying $\sum D_m = \sum \mathrm{{}^{3}He}_m = 1$. By choosing $P_\mathrm{D} P_\mathrm{{}^{3}He} = 1$, the cross section is enhanced by 50\%. Both helium-3 and deuterium nuclei must have some polarization bias for the cross section to change. While polarizing just one of deuterium or helium-3 does not change the total cross section, it does change the differential cross section \cite{Kulsrud1986}. There are also other spin-polarization schemes for D-${}^{3}$He fuel that we do not consider in this work.

In \Cref{fig:reactivities_polarized} we plot the effect of spin  polarization on the fusion reactivity for D-T, D-D, and D-${}^{3}$He versus temperature for a Maxwellian energy distribution. Solid lines correspond to unpolarized fuel. Dashed lines correspond to the expected upper bounds for the reactivity with polarized fuel. For D-T and D-${}^{3}$He, this is a 50\% enhancement relative to unpolarized fuel \cite{Kulsrud1986}, occurring when the D-T and D-${}^{3}$He nuclear spins are aligned, corresponding to $\langle \sigma_{\pm 1, \pm \frac{1}{2}} v \rangle$. The dotted lines correspond to an anti-parallel alignment of D-T and D-${}^{3}$He spins, corresponding to $\langle \sigma_{\mp 1, \pm \frac{1}{2}} v \rangle$ and a reduction in the fusion power by 50\% relative to unpolarized fuel. For D-D reactions, we take a different approach the upper and lower bounds of the reactivity, using the most optimistic and pessimistic values of the QSF for aligned D-D, since we expected in a polarized D-${}^{3}$He or D-D plasma that all the deuterium will be intended to have the same spin orientation. The upper D-D bound $\langle \sigma_{1,1}^\mathrm{T-Matrix} v \rangle$ corresponds to T-Matrix calculations \cite{Lemaitre1993} and the lower bound $\langle \sigma_{1,1}^\mathrm{T-Matrix} v \rangle$ to Alt, Grassberger, and Sandhas (AGS) calculations \cite{Deltuva2007,Engels2003}. The data was obtained from \cite{Paetz2010}. Notably for D-D, most calculations predict a strong temperature-dependence for the QSF. This is in contrast to D-T and D-${}^{3}$He, where we have assumed the cross-section enhancement/suppression to be temperature-independent.

The D-${}^{3}$He fusion power density on a flux surface with polarized fuel is
\begin{equation}
\begin{aligned}
& p_{f,\mathrm{D{}^{3}He}} = n_\mathrm{D} n_\mathrm{{}^{3}He} \left( 1 + \frac{ P_\mathrm{D} P_\mathrm{{}^{3}He}}{2} \right) \langle \overline{\sigma} v \rangle E_\mathrm{D{}^{3}He},
\end{aligned}
\label{eq:pfform2_init}
\end{equation}
where $n_\mathrm{D}$ and $n_\mathrm{{}^{3}He}$ are the deuterium and helium-3 densities and $\langle \overline{\sigma} v \rangle$ is the unpolarized Maxwellian-averaged fusion reactivity. Writing
\begin{equation}
    f_\mathrm{{}^{3}He} \equiv \frac{n_\mathrm{{}^{3}He}}{n_\mathrm{F}}, \;\;\;\; n_\mathrm{F} \equiv n_\mathrm{{}^{3}He} + n_\mathrm{D},
\end{equation}
\Cref{eq:pfform2_init} becomes
\begin{equation}
\begin{aligned}
& p_{f,\mathrm{D{}^{3}He}} = f_\mathrm{{}^{3}He} \left(1 - f_\mathrm{{}^{3}He} \right) n_\mathrm{F}^2 \left( 1 + \frac{ P_\mathrm{D} P_\mathrm{{}^{3}He}}{2} \right)  \langle \overline{\sigma} v \rangle E_\mathrm{D{}^{3}He}.
\end{aligned}
\label{eq:pfform2_init_diff}
\end{equation}
Therefore, the D-${}^{3}$He fusion power density is maximized when $f_\mathrm{{}^{3}He} = 1/2$ for constant $n_\mathrm{F}$.

However, often it is more relevant to evaluate the power density at constant electron density. Quasineutrality requires
\begin{equation}
    n_e = 2 n_\mathrm{{}^{3}He} + n_\mathrm{D} + 2 n_\mathrm{{}^{4}He} + n_\mathrm{p},
    \label{eq:quasineutrality_DHe3plasma}
\end{equation}
where the helium-4 density $n_\mathrm{{}^{4}He}$ and proton density $n_\mathrm{p}$ are sourced from fusion reactions. Defining the helium-4-to-electron density ratio and proton-to-electron density ratio \cite{Whyte2023},
\begin{equation}
    f_\mathrm{dil}^\mathrm{{}^{4}He} \equiv \frac{n_\mathrm{{}^{4}He}}{n_e}, \;\; f_\mathrm{dil}^\mathrm{p} \equiv \frac{n_\mathrm{p}}{n_e}
\end{equation}
the electron density becomes
\begin{equation}
    n_e = \frac{ 2 n_\mathrm{{}^{3}He} + n_\mathrm{D}}{1-2f_\mathrm{dil}^\mathrm{{}^{4}He} - f_\mathrm{dil}^\mathrm{p}}.
\end{equation}
The fusion power density at constant electron density is therefore
\begin{equation}
\begin{aligned}
p_{f,\mathrm{D{}^{3}He}} = & f_\mathrm{{}^{3}He} \left( 1-2f_\mathrm{dil}^\mathrm{{}^{4}He} - f_\mathrm{dil}^\mathrm{p} \right)^2 \frac{ \left(1 - f_\mathrm{{}^{3}He} \right) }{\left( 1 + f_\mathrm{{}^{3}He} \right)^2} \\
& \times n_\mathrm{e}^2 \left( 1 + \frac{ P_\mathrm{D} P_\mathrm{{}^{3}He}}{2} \right)  \langle \overline{\sigma} v \rangle E_\mathrm{D{}^{3}He}.
\end{aligned}
\label{eq:pfform3_init_diff}
\end{equation}
\Cref{eq:pfform3_init_diff} gives a maximum fusion power density at fixed electron density for $f_\mathrm{{}^{3}He} = 1/3$ with all other variables fixed -- the coupling between the fusion power density and core dilution by helium-4 and protons is similar to the coupling caused by core dilution helium-4 in D-T fusion \cite{Whyte2023,Parisi_2024d}, but has different dynamics because the proton from D-${}^{3}$He fusion does not escape the plasma as readily as the D-T neutron.

We also consider D-D and D-T fusion reactions. For each D-D reaction channel, we assume a simplified spin dependence of the fusion reactivity for polarized fuel
\begin{equation}
\langle \sigma v \rangle =
\langle \overline{\sigma} v \rangle
\Bigl[
  1 + \kappa P_1 P_2
\Bigr],
\end{equation}
where $\kappa$ is the enhancement/suppression factor and $P_1, P_2$ are the vector spin polarizations of the reactants. $P_1$ and $P_2$ can take values from -1 to +1, but for simplicity, in this paper we assume that $P_1 P_2 \geq 0$. For example, in a deuterium gas with polarization $P_\mathrm{D}$, the reactivity of the two branches is
\begin{align}
\langle \sigma v \rangle_{\mathrm{DD,n}}
&=
\langle \overline{\sigma} v \rangle_{DD,n}\;\bigl[\,1 + \kappa_{\mathrm{DD,n}} P_\mathrm{D}^2\bigr],
\label{eq:DDn_sup_enhance}
\\[6pt]
\langle \sigma v \rangle_{\mathrm{DD,p}}
&=
\langle \overline{\sigma} v \rangle_{DD,p}\;\bigl[\,1 + \kappa_{\mathrm{DD,p}} P_\mathrm{D}^2\bigr].
\label{eq:DDp_sup_enhance}
\end{align}
For the D-T reactivity, a tritium gas with polarization $P_\mathrm{T}$ and deuterium with $P_\mathrm{D}$ satisfies
\begin{equation}
    \langle \sigma v \rangle_{\mathrm{DT}} =
\langle \overline{\sigma} v \rangle_{\mathrm{DT}}
 \bigl[
  1 + \frac{P_\mathrm{D} P_\mathrm{T}}{2}
\bigr],
\end{equation}
where we use the well-known result that $\kappa = 1/2$ for D-T reactions \cite{Kulsrud1986}. Finally, using \Cref{eq:AJ2}, the D-${}^{3}$He polarized reactivity satisfies
\begin{equation}
    \langle \sigma v \rangle_{\mathrm{D{}^{3}He}} =
\langle \overline{\sigma} v \rangle_{\mathrm{D{}^{3}He}}
 \bigl[
  1 + \frac{P_\mathrm{D}\,P_\mathrm{{}^{3}He}}{2}
\bigr].
\end{equation}

\section{0D Steady-State Fusion Model with Polarized D-$^3$He Fuel} \label{sec:steadystatemodel}

In this section, we introduce the 0D system of equations to predict the power density in a plasma fueled with D-${}^{3}$He. In \Cref{subsec:mainequations} we introduce the steady state model. In \Cref{ssec:secondaryTBEHBE} we include secondary tritium and helium burn efficiencies in the model. In \Cref{ssec:temperature} we add realistic temperature dependencies to the model. In \Cref{sec:optimized_ratios} we discuss different possible optimization schemes. In \Cref{ssec:maxpower} we optimize the D-${}^{3}$He fuel mixture for maximum fusion power. In \Cref{sec:inherited_spin} we add the effect of inherited spin from D-D reactions, in \Cref{sec:branch_ratios} we discuss the impact of different reactivities for the two D-D fusion branches, and in \Cref{sec:minimize_DT_neutrons} we optimize the fuel ratio to minimize D-T neutrons.

\subsection{Steady State Model} \label{subsec:mainequations}

In this section, we describe a simplified steady state fusion power density model where as an initial assumption, all of the tritium and secondary helium-3 produced by D-D reactions is burned. We will relax this assumption later in \Cref{ssec:secondaryTBEHBE}.

The total fusion power density $P_\mathrm{f}$ is the sum of contributions from all channels,
\begin{equation}
\begin{aligned}
P_\mathrm{f}
= &
\underbrace{n_\mathrm{D}\,n_{\mathrm{{}^{3}He}}\,\langle\sigma v\rangle_{\mathrm{D{}^{3}He}} E_{\mathrm{D{}^{3}He}}}
_{\text{D-${}^{3}$He channel}}
\\
\quad
\;+ &
\underbrace{\frac12\,n_\mathrm{D}^2\,\langle\sigma v\rangle_{\mathrm{DD,n}} E_{\mathrm{DD,n}}}
_{\text{D-D (He3+n) branch}}
\\
\;+
&
\underbrace{\frac12\,n_\mathrm{D}^2\,\langle\sigma v\rangle_{\mathrm{DD,p}} E_{\mathrm{DD,p}}}
_{\text{D-D (T+p) branch}}
\\
+ &
\underbrace{n_\mathrm{D} n_{\mathrm{{}^{3}He},s}\;\langle \sigma v\rangle_{\mathrm{D{}^{3}He}} E_{\mathrm{D{}^{3}He}}}
_{\text{secondary D--$^3$He channel}}
\\
\quad
+ &
\underbrace{n_\mathrm{D}\,n_\mathrm{T}\,\langle\sigma v\rangle_{\mathrm{DT}} E_{\mathrm{DT}}}
_{\text{secondary D-T channel}}.
\end{aligned}
\label{eq:Pfusioninitial}
\end{equation}
We will keep the secondary helium density $n_{\mathrm{{}^{3}He},s}$ distinct from the main helium-3 density $n_\mathrm{{}^{3}He}$ for clarity. Our first task is to estimate the tritium and secondary helium-3 densities. Tritium is produced by $\mathrm{D}+\mathrm{D}\to \mathrm{T}+\mathrm{p}$ and consumed by $\mathrm{D}+\mathrm{T}\to\alpha+\mathrm{n}$. As an initial estimate of the tritium density, we use a zero-dimensional steady-state model with zero particle transport and losses, set the time derivative of the tritium density $n_\mathrm{T}$ to zero, and assume that all of the tritium that is produced is burned. This gives
\begin{equation}
\underbrace{
  \frac12\,n_\mathrm{D}^2\,\langle\sigma v\rangle_{\mathrm{DD,p}}
}_{\text{T production}} =
\underbrace{
  n_\mathrm{D}\,n_\mathrm{T}\,\langle\sigma v\rangle_{\mathrm{DT}}
}_{\text{T consumption}}.
\end{equation}
Therefore, the tritium density is
\begin{equation}
    n_\mathrm{T} =\frac{n_\mathrm{D}}{2}
\;\frac{\langle\sigma v\rangle_{\mathrm{DD,p}}}
      {\langle\sigma v\rangle_{\mathrm{DT}}}.
      \label{eq:nT_estimate_simple1}
\end{equation}
Note that we have neglected other tritium loss mechanisms, although we will drop this assumption in later sections. Therefore, the estimate in \Cref{eq:nT_estimate_simple1} is an upper bound on the tritium density. Inserting the steady-state $n_\mathrm{T}$ gives the fusion power density from D-T reactions
\begin{equation}
p_{\mathrm{f,DT}}
=
n_\mathrm{D}\,n_\mathrm{T} \langle\sigma v\rangle_{\mathrm{DT}} E_{\mathrm{DT}}
=
\frac{n_\mathrm{D}^2}{2} \langle\sigma v\rangle_{\mathrm{DD,p}} E_{\mathrm{DT}},
\label{eq:pfDT_simple}
\end{equation}
which (in this idealized steady-state) depends only on the T-production rate 
$\frac12\,n_\mathrm{D}^2\,\langle\sigma v\rangle_{\mathrm{DD,p}}$ and not on 
$\langle\sigma v\rangle_{\mathrm{DT}}$ itself.
This is a consequence of the assumption that T is produced and burned in a source-limited scenario.

Using a similar argument for secondary helium-3, we find an expression for the secondary helium-3 density,
\begin{equation}
    n_{\mathrm{{}^{3}He,s}} =\frac{n_\mathrm{D}}{2}
\;\frac{\langle\sigma v\rangle_{\mathrm{DD,n}}}
      {\langle\sigma v\rangle_{\mathrm{D{}^{3}He}}}.
      \label{eq:nHe3s_estimate_simple1}
\end{equation}
A similar expression to \Cref{eq:pfDT_simple} exists for the power density of secondary D${}^{3}$He reactions.

Writing the reactivity in spin-polarized form, the total steady-state fusion power density in \Cref{eq:Pfusioninitial} is
\begin{equation}
\begin{aligned}
  & p_{\mathrm{f}} = \\
  & n_\mathrm{D}\,n_\mathrm{{}^{3}He}\;\langle \overline{ \sigma} v \rangle_{\mathrm{D{}^{3}He}} \bigl[ 1 + \frac{P_\mathrm{D}\,P_\mathrm{{}^{3}He}}{2} \bigr] E_{\mathrm{D{}^{3}He}}
  \;+\; \\
  & \tfrac12\,n_\mathrm{D}^2 \langle \overline{\sigma} v \rangle_{\mathrm{DD,n}} \bigl[1 + \kappa_{\mathrm{DD,n}} P_\mathrm{D}^2\bigr] \left( E_{\mathrm{DD,n}} + E_{\mathrm{D{}^{3}He}} \right)
   + \\
  & \tfrac12\,n_\mathrm{D}^2 \langle \overline{\sigma} v \rangle_{\mathrm{DD,p}}\,\bigl[ 1 + \kappa_{\mathrm{DD,p}} P_\mathrm{D}^2\bigr]
  \;\bigl(E_{\mathrm{DD,p}} + E_{\mathrm{DT}}\bigr).
\end{aligned}
\label{eq:Pfusion_earlySPF}
\end{equation}

\begin{figure*}[tb!]
    \centering
    \begin{subfigure}[t]{0.49\textwidth}
    \centering
    \includegraphics[width=1.0\textwidth]{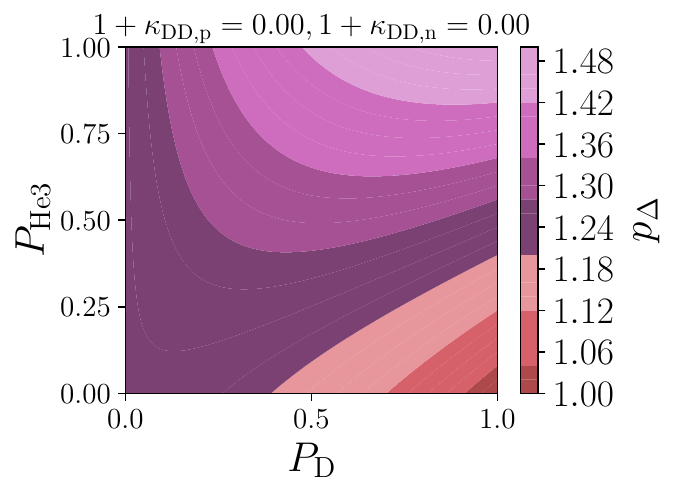}
    \caption{}
    \end{subfigure}
    \centering
    \begin{subfigure}[t]{0.49\textwidth}
    \centering
    \includegraphics[width=1.0\textwidth]{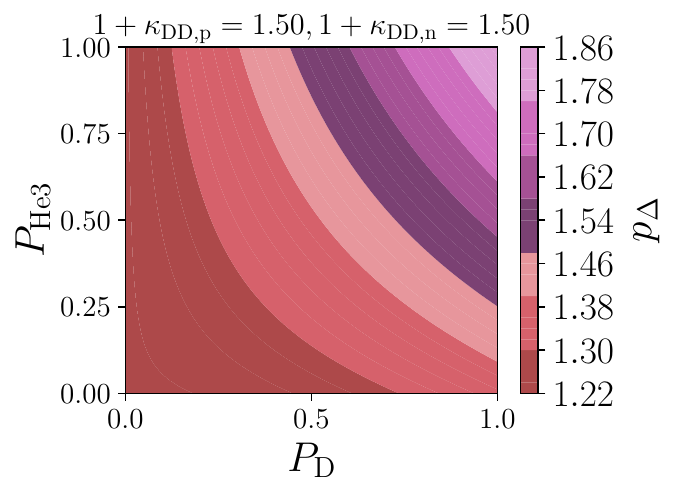}
    \caption{}
    \end{subfigure}
    \caption{Fusion power multiplier $p_\Delta$ (\Cref{eq:pDeltaform1}) for deuterium and helium-3 vector polarization $P_\mathrm{D}$ and $P_\mathrm{{}^{3}He}$. (a) $1 + \kappa_{\mathrm{DD,p}} = 0.0$ and $1 + \kappa_{\mathrm{DD,n}} = 0.0$, (b) $1 + \kappa_{\mathrm{DD,p}} = 1.5$ and $1 + \kappa_{\mathrm{DD,n}} = 1.5$ values (see \Cref{eq:DDn_sup_enhance,eq:DDp_sup_enhance}). We have assumed a plasma temperature of 50 keV, have ignored all loss mechanisms, and used $n_\mathrm{{}^{3}He} = n_\mathrm{D}$.}
    \label{fig:pDelta_DeHe3}
\end{figure*}

\noindent
Notice that $\langle\sigma v\rangle_{\mathrm{DT}}$ (and its spin factor $\kappa_{\mathrm{DT}}$) does not appear, because 
the D-T channel becomes source-limited under these idealized assumptions. The increase in fusion power due to side-reactions and the spin-polarization relative to unpolarized D-${}^{3}$He power is
\begin{equation}
    \begin{aligned}
        & p_\Delta \equiv \frac{p_\mathrm{f}}{p_{\mathrm{f,0}}} =  1 + \\
        & \frac{n_\mathrm{D}}{2 n_\mathrm{{}^{3}He}} \frac{\langle \overline{\sigma} v \rangle_{\mathrm{DD,n}}}{\langle \overline{ \sigma} v \rangle_{\mathrm{D{}^{3}He}}} \frac{\,\bigl[\,1 + \kappa_{\mathrm{DD,n}}\,P_\mathrm{D}^2\bigr]\; \left(E_{\mathrm{DD,n}} + E_{\mathrm{D{}^{3}He}} \right)}{E_{\mathrm{D{}^{3}He}}} \\
        & + \frac{n_\mathrm{D}}{2 n_\mathrm{{}^{3}He}} \frac{\langle \overline{\sigma} v \rangle_{\mathrm{DD,p}}}{\langle \overline{ \sigma} v \rangle_{\mathrm{D{}^{3}He}}} \frac{\bigl[\,1 + \kappa_{\mathrm{DD,p}}\,P_\mathrm{D}^2\bigr]
  \;\bigl(E_{\mathrm{DD,p}} + E_{\mathrm{DT}}\bigr)}{E_{\mathrm{D{}^{3}He}}},
    \end{aligned}
    \label{eq:pDeltaform1}
\end{equation}
where $p_{\mathrm{f,0}}$ is the fusion power from unpolarized D-${}^3$He reactions. In \Cref{fig:pDelta_DeHe3} we plot $p_\Delta$ for different suppression / enhancement factors of the D-D branches, corresponding to $1 + \kappa_{\mathrm{DD,n}}$ and $1 + \kappa_{\mathrm{DD,p}}$. Because we have included multiple fusion reaction channels, the power enhancement/suppression does not follow contours of constant $P_\mathrm{D} P_\mathrm{{}^3He}$ as it would if we only considered D-${}^3$He reactions (\Cref{eq:pfform2_init}). In \Cref{fig:pDelta_DeHe3}(a), $\kappa_{\mathrm{DD,n}} = \kappa_{\mathrm{DD,p}} = -1$, which means D-D reactions are fully suppressed when $P_\mathrm{D} = 1$ -- this is why $p_\Delta = 1$ for $P_\mathrm{D} = 1$, $P_\mathrm{{}^3He} = 0$. In \Cref{fig:pDelta_DeHe3}(b), $\kappa_{\mathrm{DD,n}} = \kappa_{\mathrm{DD,p}} = 0.5$, which causes $p_\Delta$ to increase monotonically with both $P_\mathrm{D}$ and $P_\mathrm{{}^3He}$ -- when $P_\mathrm{D} = P_\mathrm{{}^3He} = 1$, the fusion power is 86\% higher than for unpolarized D-${}^3$He reactions.

\subsection{Secondary Tritium and Helium Burn Efficiency} \label{ssec:secondaryTBEHBE}

In this section, we relax the assumption that all of the tritium and secondary helium-3 is burned by introducing a secondary burn efficiency -- the fraction of tritium or secondary helium-3 burned in fusion reactions.

It is important to note that the fusion reactivities for D-T and secondary D-${}^{3}$He reactions are evaluated at much higher temperature than the D-D and primary D-${}^{3}$He reactions -- this is because the tritium and secondary helium-3 are born at a minimum of 1.0 MeV and 0.8 MeV respectively. For thermal distributions, the D-T reactivity peaks at $\sim$60 keV (see \Cref{fig:reactivities_polarized}). As long as the plasma deuterium temperature is below the birth energy of 1.0 MeV, the D-T reactivity for a tritium particle slowing down in the plasma through collisions will increase until the tritium temperature drops below $\sim$60 keV. A similar phenomenon exists for secondary helium-3 since the peak D-${}^{3}$He reactivity occurs at $\sim$250 keV. This is a curious form of plasma heating where the reactivity of the D-D-born tritium and secondary helium-3 initially increases while they slow down on the plasma through collisions, assuming that the tritum and helium-3 are well-confined as they slow in the plasma. An important result of this argument is that the fusion reactivity of secondary D-${}^{3}$He reactions $\langle\sigma v\rangle_{\mathrm{D{}^{3}He,s}}$ is in general not equal to the primary D-${}^{3}$He reactivity,
\begin{equation}
    \langle\sigma v\rangle_{\mathrm{D{}^{3}He,s}} \neq \langle\sigma v\rangle_{\mathrm{D{}^{3}He}}.
\end{equation}
However, it is important to note that the D-${}^{3}$He reactivity is fairly independent of temperature between 0.1 and 1.0 MeV, varying at most by a factor of $\sim$2 (\Cref{fig:reactivities_polarized}) -- therefore, for the higher temperatures generally desired for D-${}^{3}$He fusion, $\langle\sigma v\rangle_{\mathrm{D{}^{3}He,s}}$ and $\langle\sigma v\rangle_{\mathrm{D{}^{3}He}}$ may not differ by more than a factor of $\sim$2 unless the equilibrium deuterium temperature is lower than 0.1 MeV. This argument is an oversimplification because the reactivities evaluated in \Cref{fig:reactivities_polarized} are for thermal distributions -- we expect non-thermal secondary helium-3 and tritium distributions (which can enhance or suppress reaction rates \cite{Hay_2015,putvinski2019fusion,kong2023enhancement,squarer2024enhancement}) if the secondary burn-up fraction is high.

\begin{figure*}[tb!]
    \centering
    \begin{subfigure}[t]{1.\textwidth}
    \centering
    \includegraphics[width=1.0\textwidth]{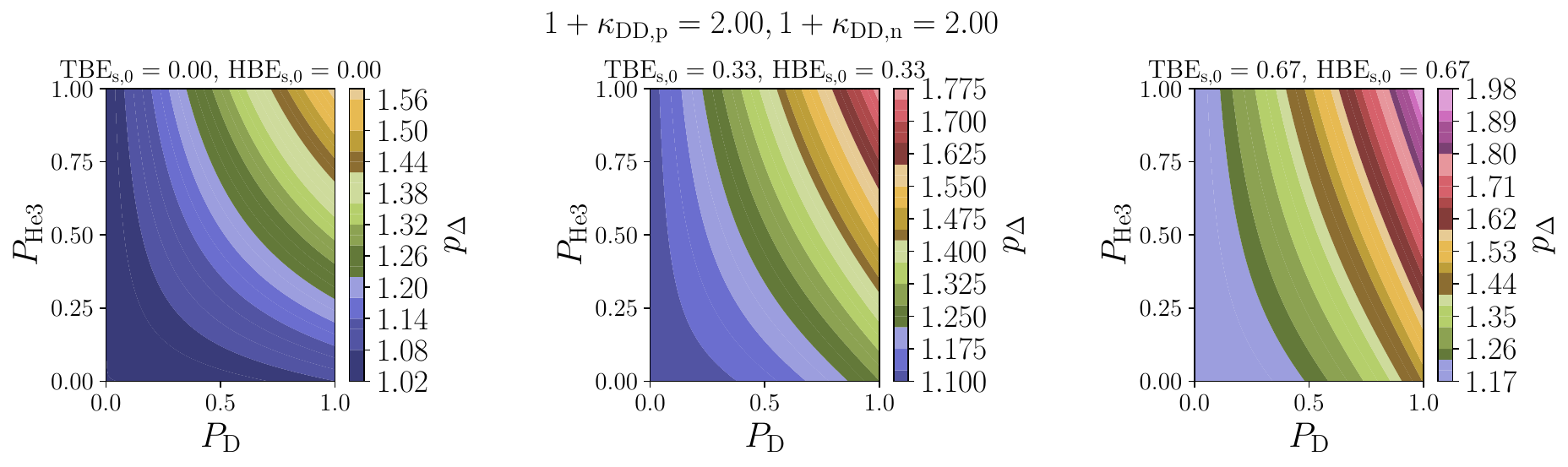}
    \end{subfigure}
    \caption{Fusion power multiplier $p_\Delta$ (\Cref{eq:pDeltaform1}) for deuterium and helium-3 vector polarization $P_\mathrm{D}$ and $P_\mathrm{{}^{3}He}$. Each subplot corresponds to a different $\mathrm{TBE}_\mathrm{s,0}$ and $\mathrm{HBE}_\mathrm{s,0}$ value. We assume that the enhancement / suppression factor of the D-D fusion reaction satisfies $1 + \kappa_{\mathrm{DD,p}} = 2$ and $1 + \kappa_{\mathrm{DD,n}} = 2$ (see \Cref{eq:DDn_sup_enhance,eq:DDp_sup_enhance}), a plasma temperature of 50 keV, and equal helium-3 and deuterium density, $n_\mathrm{{}^{3}He} = n_\mathrm{D}$.}
    \label{fig:pDelta_DeHe3TBE}
\end{figure*}

For tritium, the instantaneous secondary burn efficiency is
\begin{equation}
    \mathrm{TBE}_\mathrm{s} \equiv \frac{\dot{n}_\mathrm{T}^\mathrm{burn}}{\dot{n}^\mathrm{source}_\mathrm{T}} = 2 \frac{n_\mathrm{T}}{n_\mathrm{D}} \frac{\langle\sigma v\rangle_{\mathrm{DT}}}{\langle\sigma v\rangle_{\mathrm{DD,p}}},
    \label{eq:TBE2}
\end{equation}
where
\begin{equation}
    \dot{n}_\mathrm{T}^\mathrm{burn} = n_\mathrm{D} n_\mathrm{T} \langle\sigma v\rangle_{\mathrm{DT}},
\end{equation}
and
\begin{equation}
    \dot{n}_\mathrm{T}^\mathrm{source} = \frac12 n_\mathrm{D}^2 \langle\sigma v\rangle_{\mathrm{DD,p}}.
\end{equation}
In the steady state model in \Cref{subsec:mainequations}, we had assumed $\mathrm{TBE}_\mathrm{s} = 1$. Now we drop the assumption of steady state ($\dot{n}_\mathrm{T}$ = 0) and / or no transport, and we assume that the tritium production rate is balanced by tritium burn and transport $\dot{n}_\mathrm{T}^\Gamma$,
\begin{equation}
    \dot{n}_\mathrm{T}^\mathrm{source} = \dot{n}_\mathrm{T}^\mathrm{burn} + \dot{n}_\mathrm{T}^\Gamma + \dot{n}_\mathrm{T}.
    \label{eq:particlebalance}
\end{equation}
In a pulsed device, we simplify the burn phase by writing the transport term as
\begin{equation}
    \dot{n}_\mathrm{T}^\Gamma = L v_\mathrm{T}^\mathrm{c} n_\mathrm{T},
\end{equation}
where $L$ is a geometric term with dimensions of inverse length and $v_\mathrm{T}^\mathrm{c}$ is a convective tritium velocity. Substituting this into \Cref{eq:particlebalance,eq:TBE2} gives
\begin{equation}
    \mathrm{TBE}_\mathrm{s} = 1 - 2  \frac{n_\mathrm{T}}{n_\mathrm{D}} \frac{1}{n_\mathrm{D}\,\langle\sigma v\rangle_{\mathrm{DD,p}}} \left[ L v_\mathrm{T}^\mathrm{c} + \frac{ \dot{n}_\mathrm{T}}{n_\mathrm{T}} \right].
\end{equation}
Similarly, the secondary helium burn efficiency is,
\begin{equation}
    \mathrm{HBE}_\mathrm{s} \equiv \frac{\dot{n}_\mathrm{He3,s}^\mathrm{burn}}{\dot{n}^\mathrm{source}_\mathrm{{}^{3}He}} = 2 \frac{n_\mathrm{{}^3He,s}}{n_\mathrm{D}} \frac{\langle\sigma v\rangle_{\mathrm{D{}^{3}He}}}{\langle\sigma v\rangle_{\mathrm{DD,n}}}.
    \label{eq:TBE2}
\end{equation}
The helium-3 density resulting from D-D production satisfies the secondary helium-3 transport equation
\begin{equation}
    \dot{n}_\mathrm{He3,s}^\mathrm{source} = \dot{n}_\mathrm{He3,s}^\mathrm{burn} + \dot{n}_\mathrm{He3,s}^\Gamma + \dot{n}_\mathrm{He3,s}.
    \label{eq:particlebalanceHe3s}
\end{equation}
Assuming we know the secondary tritium and helium burn efficiencies, the fusion power is a modified version of \Cref{eq:Pfusion_earlySPF},
\begin{equation}
\begin{aligned}
  & p_{\mathrm{f}} = \\
  & \tfrac12\,n_\mathrm{D}^2 \langle \overline{\sigma} v \rangle_{\mathrm{DD,n}} \bigl[ 1 + \kappa_{\mathrm{DD,n}} P_\mathrm{D}^2\bigr] \left( E_{\mathrm{DD,n}} + \mathrm{HBE}_\mathrm{s} E_{\mathrm{D{}^{3}He}} \right)
  \;+\; \\
  & \tfrac12\,n_\mathrm{D}^2\;\langle \overline{\sigma} v \rangle_{\mathrm{DD,p}}\,\bigl[\,1 + \kappa_{\mathrm{DD,p}}\,P_\mathrm{D}^2\bigr]
  \;\bigl(E_{\mathrm{DD,p}} + \mathrm{TBE}_\mathrm{s} E_{\mathrm{DT}}\bigr) \;+\; \\
  & n_\mathrm{D}\,n_{\mathrm{{}^{3}He},0}\;\langle \overline{ \sigma} v \rangle_{\mathrm{D{}^{3}He}} \bigl[1 + \frac{P_\mathrm{D} P_\mathrm{{}^{3}He}}{2} \bigr] E_{\mathrm{D{}^{3}He}}.
\end{aligned}
\label{eq:Pfusion_secondary_burn_efficiency}
\end{equation}
The increase in fusion power due to side-reactions and the spin-polarization relative to unpolarized D-${}^{3}$He power is
\begin{equation}
    \begin{aligned}
        & p_\Delta = 1 + \\
        & \frac{n_\mathrm{D}}{2 n_{\mathrm{{}^{3}He}}} \frac{\langle \overline{\sigma} v \rangle_{\mathrm{DD,n}}}{\langle \overline{ \sigma} v \rangle_{\mathrm{D{}^{3}He}}} \frac{\,\bigl[\,1 + \kappa_{\mathrm{DD,n}}\,P_\mathrm{D}^2\bigr]\; \left(E_{\mathrm{DD,n}} +  \mathrm{HBE}_\mathrm{s} E_{\mathrm{D{}^{3}He}} \right)}{E_{\mathrm{D{}^{3}He}}} \\
        & + \frac{n_\mathrm{D}}{2 n_{\mathrm{{}^{3}He}}} \frac{\langle \overline{\sigma} v \rangle_{\mathrm{DD,p}}}{\langle \overline{ \sigma} v \rangle_{\mathrm{D{}^{3}He}}} \frac{\bigl[\,1 + \kappa_{\mathrm{DD,p}}\,P_\mathrm{D}^2\bigr]
  \;\bigl(E_{\mathrm{DD,p}} +  \mathrm{TBE}_\mathrm{s} E_{\mathrm{DT}}\bigr)}{E_{\mathrm{D{}^{3}He}}}.
    \end{aligned}
    \label{eq:pDeltaform2}
\end{equation}

One might expect $\mathrm{HBE}_\mathrm{s}$ and $\mathrm{TBE}_\mathrm{s}$ depend on polarization because cross section increases with polarization. To that end, we implement a polarization-dependent burn efficiency model
\begin{equation}
\begin{aligned}
    & \mathrm{HBE}_\mathrm{s} = \mathrm{HBE}_\mathrm{s,0} \left( 1 + \frac{P_\mathrm{D} P_\mathrm{{}^{3}He,s}}{2} \right), \\
    & \mathrm{TBE}_\mathrm{s} = \mathrm{TBE}_\mathrm{s,0} \left( 1 + \frac{P_\mathrm{D} P_\mathrm{T}}{2} \right),
\end{aligned}
\label{eq:burn_efficiencies}
\end{equation}
where $\mathrm{HBE}_\mathrm{s,0}$ and $\mathrm{TBE}_\mathrm{s,0}$ are the burn efficiencies with zero polarization. For now, we assume that the secondary helium-3 and tritium spin polarizations are fully inherited from the deuterium spin polarization: $P_\mathrm{{}^{3}He,s} = P_\mathrm{T} = P_\mathrm{D}$. We will relax this assumption later in \Cref{sec:inherited_spin}.

In \Cref{fig:pDelta_DeHe3TBE}, we plot $p_\Delta$ for deuterium and helium-3 vector polarization $P_\mathrm{D}$ and $P_\mathrm{{}^{3}He}$ using the burn efficiency model in \Cref{eq:burn_efficiencies}. Each subplot corresponds to three different $\mathrm{TBE}_\mathrm{s,0}$ and $\mathrm{HBE}_\mathrm{s,0}$ values -- higher $\mathrm{TBE}_\mathrm{s,0}$ and $\mathrm{HBE}_\mathrm{s,0}$ lead to higher $p_\Delta$ values. In \Cref{eq:burn_efficiencies}(c) for $P_\mathrm{D} P_\mathrm{{}^{3}He} = 1$, roughly 25\% of the fusion power comes from D-D and secondary reactions. Inspecting \Cref{eq:pDeltaform2}, we see that secondary D-${}^{3}$He power exceeds the primary D-D neutronic branch when
\begin{equation}
    \mathrm{HBE}_\mathrm{s} > \frac{E_{\mathrm{DD,n}}}{ E_{\mathrm{D{}^{3}He}}} \simeq 0.18,
\end{equation}
and secondary D-T power exceeds the primary D-D aneutronic branch when
\begin{equation}
    \mathrm{TBE}_\mathrm{s} > \frac{E_{\mathrm{DD,p}}}{ E_{\mathrm{DT}}} \simeq 0.23.
\end{equation}
Given that the tritium and helium-3 produced from D-D reactions are at MeV energies, such a high burn efficiency might be attainable given the much higher cross section. Indeed, one might also expect that $\mathrm{HBE}_\mathrm{s} \approx \mathrm{TBE}_\mathrm{s}$ because the D-${}^{3}$He and D-T fusion cross sections are comparable at MeV energies.

\subsection{Temperature Dependence} \label{ssec:temperature}

\begin{figure}[tb!]
    \centering
    \begin{subfigure}[t]{0.97\textwidth}
    \centering
    \includegraphics[width=1.0\textwidth]{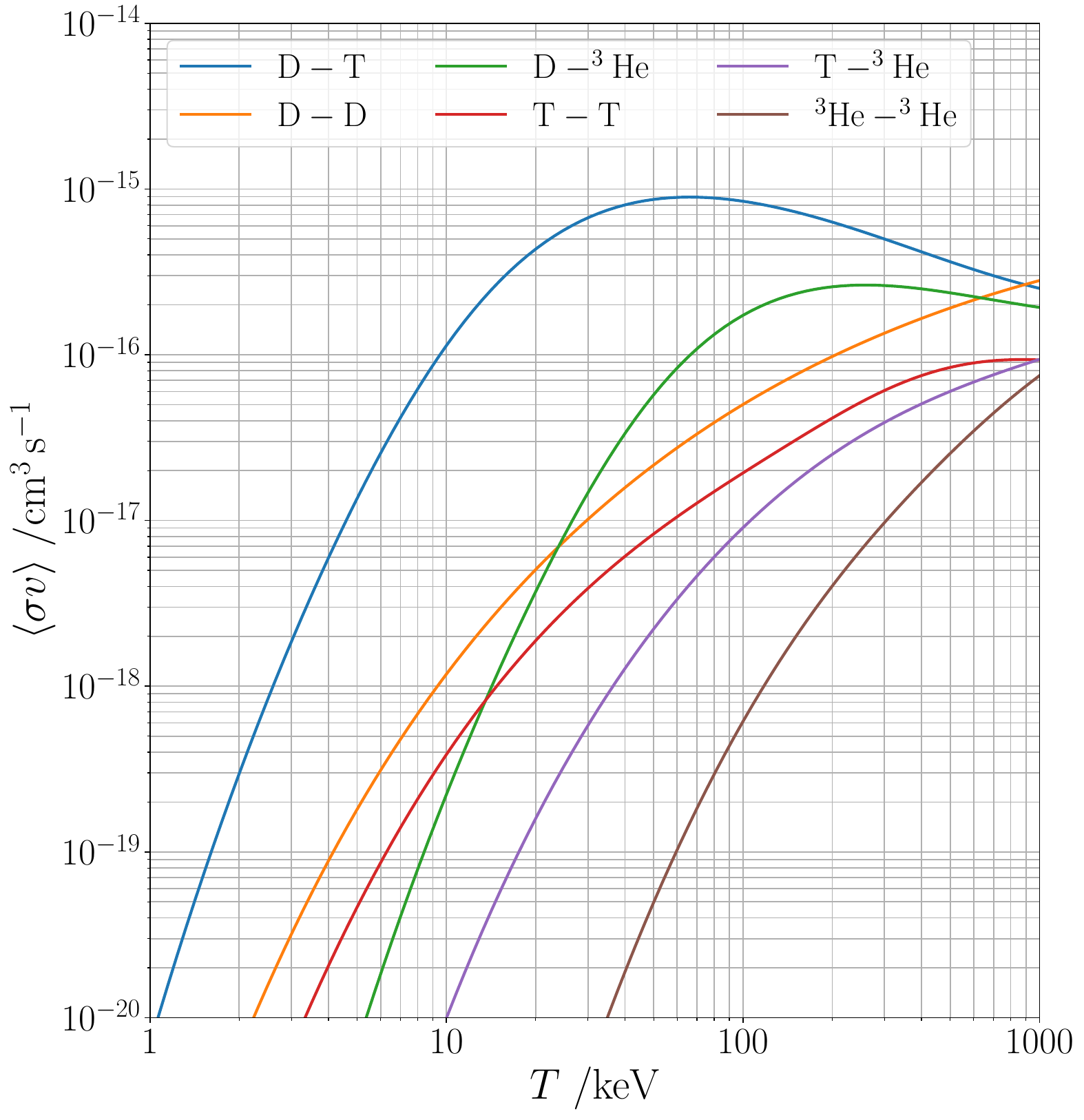}
    \end{subfigure}
    \caption{Unpolarized reactivities for some fusion fuels with a Maxwellian energy distribution.}
    \label{fig:reactivities}
\end{figure}

Up to now, we have produced results with a constant temperature. However, as temperature increases the ratios of the reactivities for different fusion reactions change significantly. As a reminder, in \Cref{fig:reactivities} we plot the standard unpolarized reactivities for some standard fusion reactions.

\begin{figure*}[tb!]
    \centering
    \begin{subfigure}[t]{1.\textwidth}
    \centering
    \includegraphics[width=1.0\textwidth]{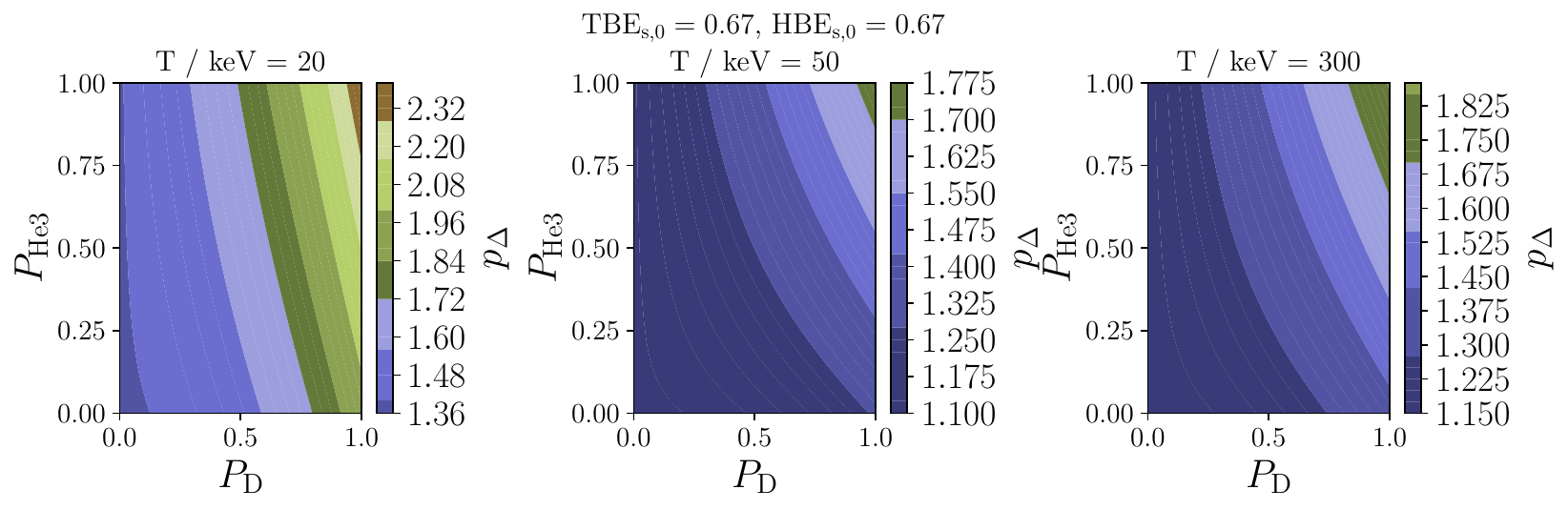}
    \end{subfigure}
    \centering
    \begin{subfigure}[t]{1.\textwidth}
    \centering
    \includegraphics[width=1.0\textwidth]{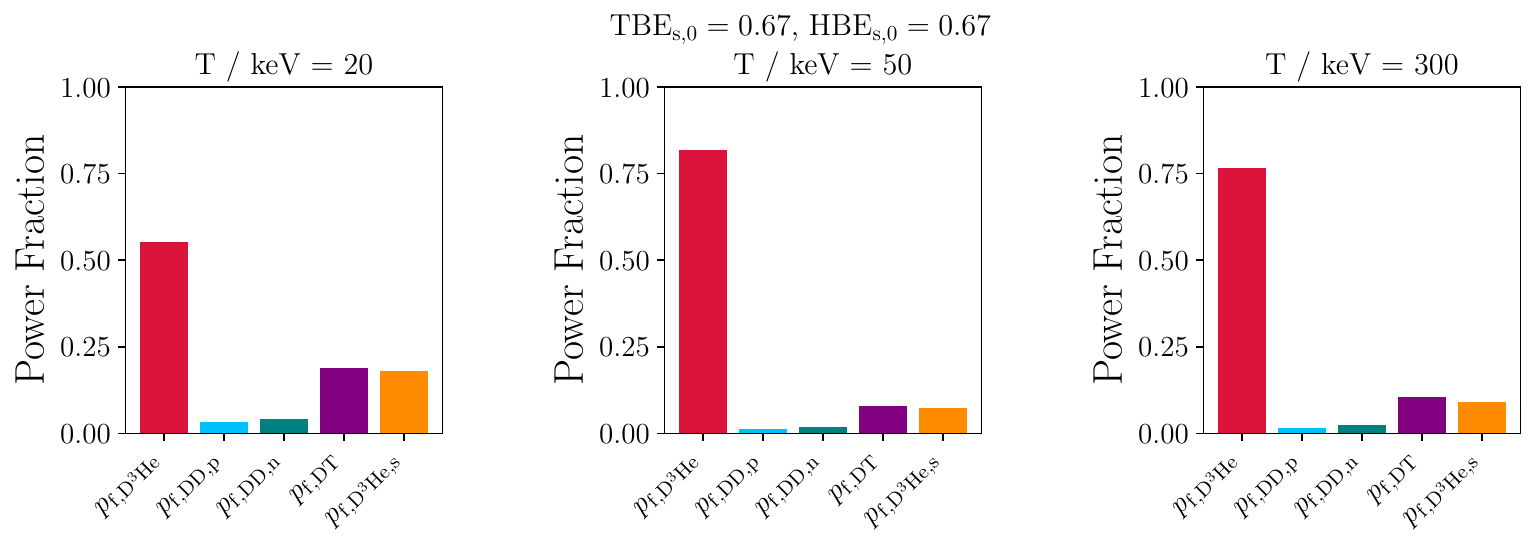}
    \end{subfigure} 
    \caption{Top: Fusion power multiplier $p_\Delta$ (\Cref{eq:pDeltaform1}) for deuterium and helium-3 vector polarization $P_\mathrm{D}$ and $P_\mathrm{{}^{3}He}$. Each subplot corresponds to a different temperature. We assume $1 + \kappa_{\mathrm{DD,p}} = 1.5$ and $1 + \kappa_{\mathrm{DD,n}} = 1.5$, $\mathrm{TBE}_\mathrm{s,0} = \mathrm{HBE}_\mathrm{s,0} = 0.67$, and $n_\mathrm{{}^{3}He} = n_\mathrm{D}$. Bottom: fraction of fusion power coming from each reaction for the $P_\mathrm{D} = P_\mathrm{{}^{3}He} = 1.0$ case. }
    \label{fig:temperaturedependent}
\end{figure*}

We now perform the previous exercise but with a temperature-dependent scan. In the top 3 plots of \Cref{fig:temperaturedependent}, we show $p_\Delta$ and in the bottom 3 plots we show the corresponding power fraction from each fusion channel for $P_\mathrm{D} = P_\mathrm{{}^{3}He} = 1.0$. At lower temperature roughly half of the fusion power comes from D-${}^3$He reactions and half from D-D reactions and secondary reactions. As the temperature increases more fusion power comes from D-${}^3$He reactions because the reactivity is relatively higher (see \Cref{fig:reactivities}).

\subsection{Optimized Fuel Ratios and Polarization} \label{sec:optimized_ratios}

Up until now, we have assumed that the deuterium and helium densities are equal,
\begin{equation}
    n_\mathrm{D} = n_\mathrm{{}^{3}He}.
\end{equation}
However, because of D-D reactions and subsequent secondary reactions, $n_\mathrm{D} = n_\mathrm{{}^{3}He}$ is far from the optimal fuel mix for maximum fusion power. In the limit of zero dilution from helium-4 and protons, \Cref{eq:pfform3_init_diff} shows how $n_\mathrm{{}^{3}He} = n_\mathrm{D}/2$ gives the maximum power density for D-${}^{3}$He fusion. In the next section, we find the optimal $n_\mathrm{{}^{3}He}$ to $n_\mathrm{D}$ fuel ratio considering all significant fusion reactions in a D-${}^{3}$He plasma.

In the subsequent sections, we discuss strategies for optimizing the D-${}^{3}$He fuel ratio for different purposes. There may be several competing objectives:
\begin{enumerate}
    \item Maximize fusion power
    \item Minimize D-T neutrons
    \item Helium-3 self-sufficiency
\end{enumerate}
When optimizing for each objective, we will keep the electron density constant.

\subsection{Optimized Fuel Ratio: Maximize fusion power} \label{ssec:maxpower}

In order to find the optimized fuel ratio, we first define $\alpha$ to represent the fraction of electrons that come from $^3$He,
\begin{equation}\label{eq:alpha}
\alpha = \frac{2 n_\mathrm{{}^{3}He}}{n_e}.
\end{equation}
Our goal is to find the $\alpha^*$ that maximizes $p_\mathrm{f}(\alpha)$,
\begin{equation}
\frac{d}{d\alpha} p_\mathrm{f}(\alpha) \bigg{|}_{\alpha^*} = 0.
\end{equation}
Quasineutrality for a D-${}^3$He plasma with D-${}^3$He, D-D, and secondary D-${}^3$He and D-T reactions is
\begin{equation}
\begin{aligned}
    n_e = & 2 n_\mathrm{{}^{3}He} + n_\mathrm{D} + 2 n_\mathrm{{}^{4}He} + n_\mathrm{p} \\
    + & 2 n_\mathrm{{}^{3}He,s} + 2  n_\mathrm{{}^{4}He,s (D-{}^3He)} + n_\mathrm{p,s} \\
    + & n_{\mathrm{T}} + 2 n_\mathrm{{}^{4}He,s (D-T)},
    \label{eq:quasineutrality_DHe3plasma_secondaries}
\end{aligned}
\end{equation}
where $n_\mathrm{{}^{4}He,s (D-{}^3He)}$ is the helium-4 density arising from secondary D-${}^3$He reactions, $n_\mathrm{{}^{4}He,s (D-T)}$ is the helium-4 density arising from D-T reactions, and $n_\mathrm{p,s}$ is the proton density arising from secondary D-${}^3$He reactions.

\begin{figure}[tb!]
    \centering
    \begin{subfigure}[t]{0.97\textwidth}
    \centering
    \includegraphics[width=1.0\textwidth]{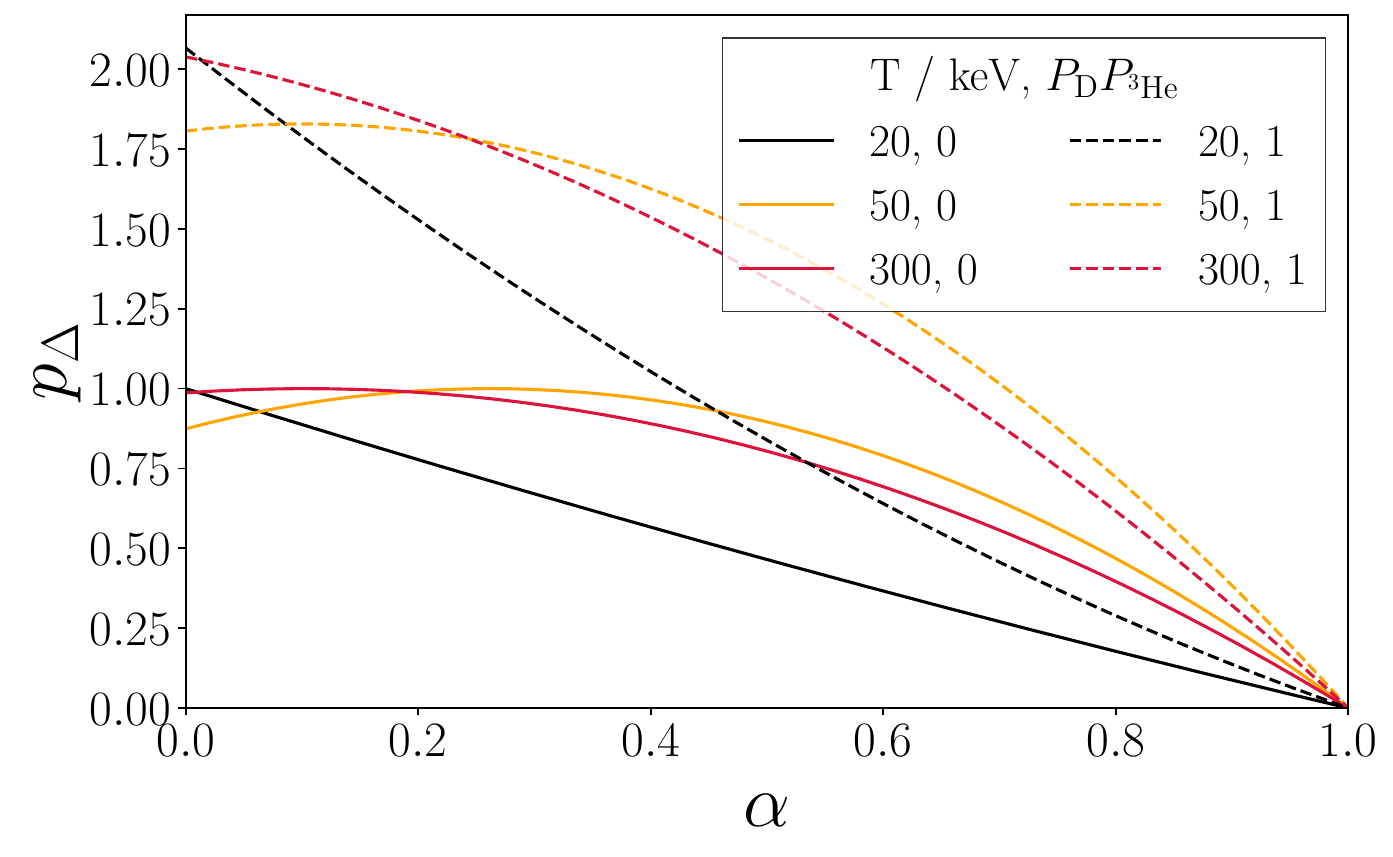}
    \end{subfigure}
    \caption{Power density enhancement/suppression $p_\Delta$ in \Cref{eq:pf_function_of_alpha} versus $\alpha$ for three temperatures and unpolarized and fully polarized fuel -- for each curve, $p_\mathrm{f}$ is normalized to the maximum power density (in $\alpha$) for unpolarized fuel at the same temperature.}
    \label{fig:pDelta_versus_alpha}
\end{figure}

Estimating the densities of all the different species in \Cref{eq:quasineutrality_DHe3plasma_secondaries} is a task beyond the scope of this work. It is challenging -- but not intractable -- because it involves significant algebra and many assumptions to make transport models for the different species. Because implementing an accurate model would involve a serious undertaking, introducing additional uncertainties of its own, we decide to make the simplest-possible model and neglect all ion species in \Cref{eq:quasineutrality_DHe3plasma_secondaries} except for the two initial fuel species, deuterium and helium-3,
\begin{equation}
    n_e = 2 n_\mathrm{{}^{3}He} + n_\mathrm{D}.
    \label{eq:DHe3_quasineutrality_simps}
\end{equation}
Taking \Cref{eq:Pfusion_secondary_burn_efficiency}, substituting $\alpha$, and using quasineutrality in \Cref{eq:DHe3_quasineutrality_simps} gives the power density
\begin{equation}
\begin{aligned}
  & \frac{p_{\mathrm{f}}}{n_\mathrm{e}^2} = \alpha (1- \alpha) \langle \overline{ \sigma} v \rangle_{\mathrm{D{}^{3}He}} \bigl[1 + \frac{P_\mathrm{D}P_\mathrm{{}^{3}He}}{2} \bigr] E_{\mathrm{D{}^{3}He}}
  + \\
  & \frac{ (1- \alpha)^2}{2} \langle \overline{\sigma} v \rangle_{\mathrm{DD,p}}\bigl[1 + \kappa_{\mathrm{DD,p}}P_\mathrm{D}^2\bigr]
  \bigl(E_{\mathrm{DD,p}} + \mathrm{TBE}_\mathrm{s} E_{\mathrm{DT}}\bigr) + \\
  & \frac{ (1- \alpha)^2}{2} \langle \overline{\sigma} v \rangle_{\mathrm{DD,n}}\bigl[1 + \kappa_{\mathrm{DD,n}}P_\mathrm{D}^2\bigr] \left( E_{\mathrm{DD,n}} + \mathrm{HBE}_\mathrm{s} E_{\mathrm{D{}^{3}He}} \right).
\end{aligned}
\label{eq:pf_function_of_alpha}
\end{equation}
In \Cref{fig:pDelta_versus_alpha}, we plot solutions to the power density enhancement/suppression $p_\Delta$ in \Cref{eq:pf_function_of_alpha} versus $\alpha$ -- for each curve, $p_\mathrm{f}$ is normalized to the maximum power density (in $\alpha$) for unpolarized fuel at the same temperature.

\begin{figure*}[tb!]
    \centering
    \begin{subfigure}[t]{1.\textwidth}
    \centering
    \includegraphics[width=1.0\textwidth]{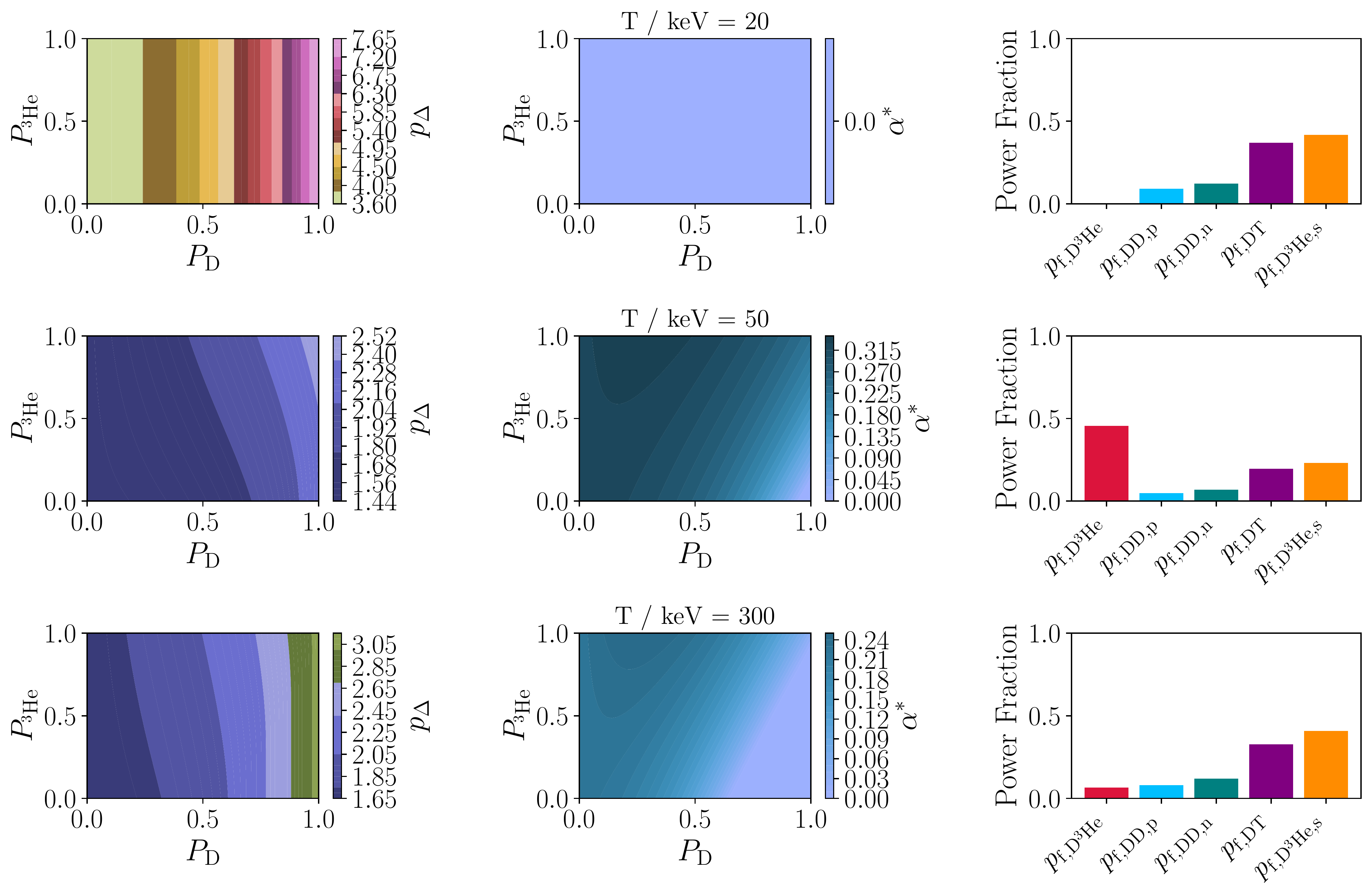}
    \end{subfigure}
    \caption{D-${}^{3}$He system with D and ${}^{3}$He fuel ratio optimized for maximum fusion power. Left column: Fusion power multiplier $p_\Delta$ (\Cref{eq:pDeltaform1}) for deuterium and helium-3 vector polarization $P_\mathrm{D}$ and $P_\mathrm{{}^{3}He}$. Middle column: optimal $\alpha = 2n_\mathrm{{}^{3}He}/n_e$ value for maximizing the fusion power. Right column: fraction of fusion power coming from each reaction for the $P_\mathrm{D} = P_\mathrm{{}^{3}He} = 1.0$ case. Each row corresponds to a different temperature. Here $1 + \kappa_{\mathrm{DD,p}} = 1.5$ and $1 + \kappa_{\mathrm{DD,n}} = 1.5$ and $\mathrm{TBE}_\mathrm{s} = \mathrm{HBE}_\mathrm{s} = 0.50$. }
    \label{fig:temperaturedependent_optimized_alpha}
\end{figure*}

\begin{figure}[tb!]
    \centering
    \begin{subfigure}[t]{0.99\textwidth}
    \centering
    \includegraphics[width=1.0\textwidth]{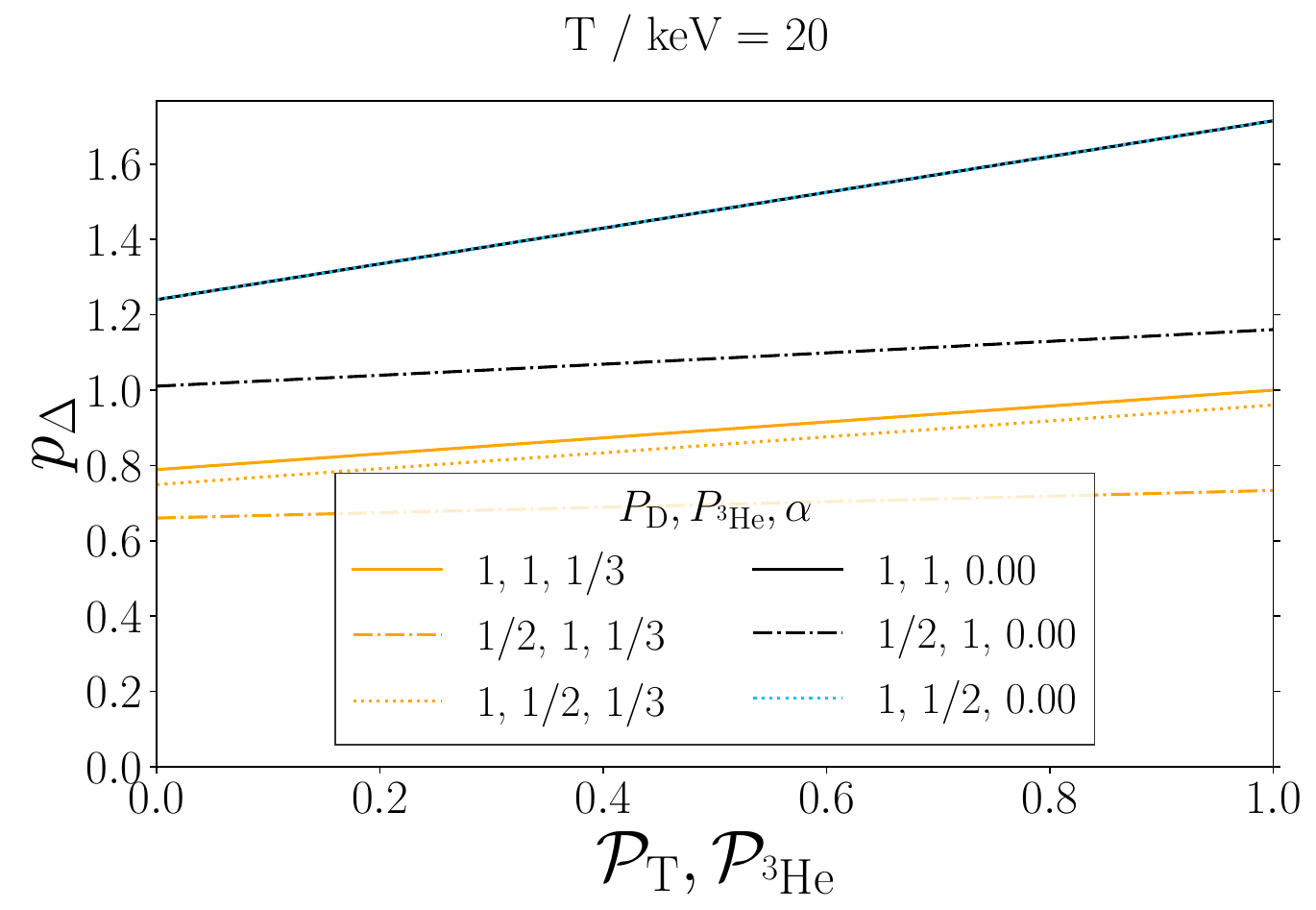}
    \caption{}
    \end{subfigure}
    \centering
    \begin{subfigure}[t]{0.99\textwidth}
    \centering
    \includegraphics[width=1.0\textwidth]{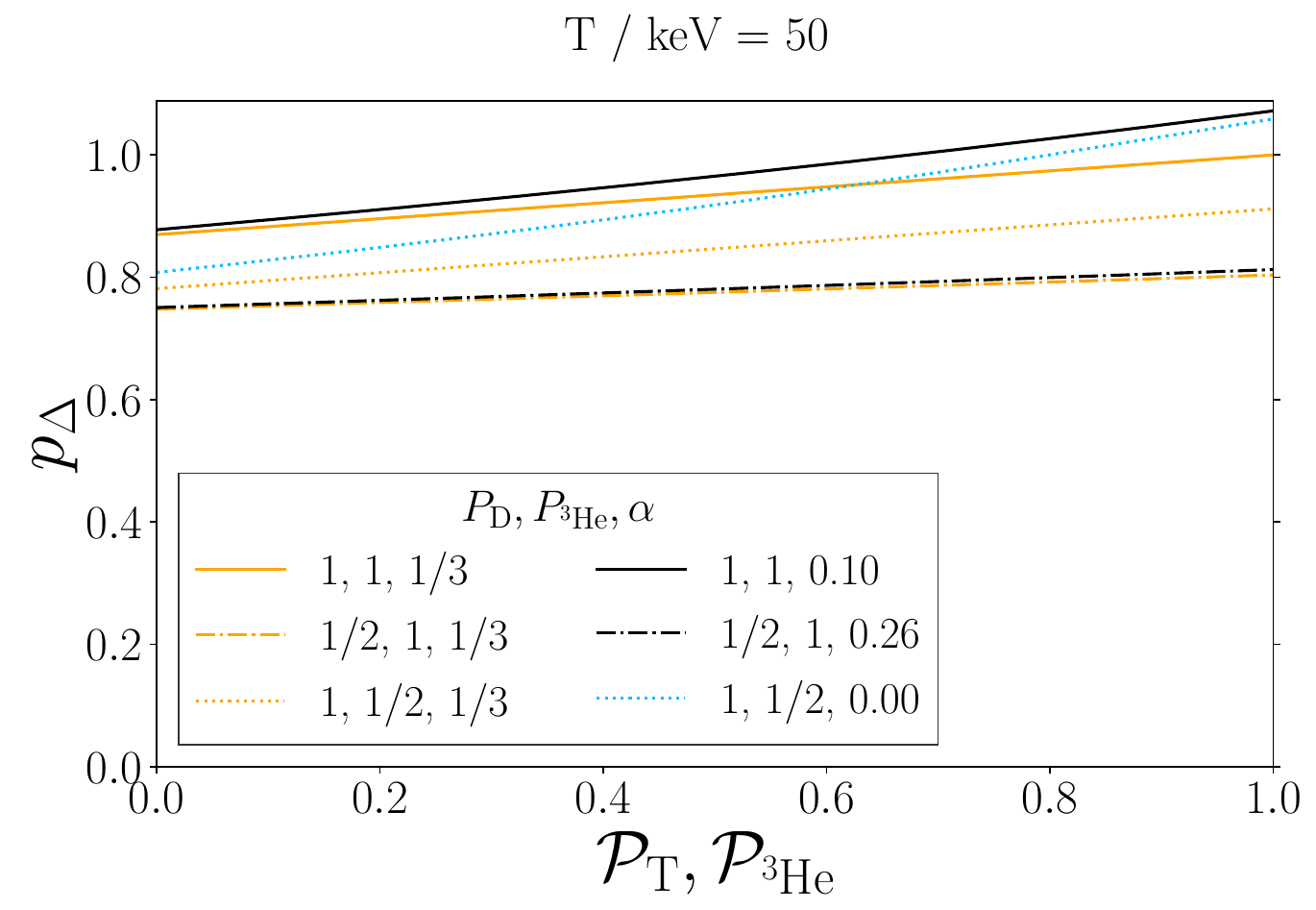}
    \caption{}
    \end{subfigure}
    \caption{Power density enhancement/suppression $p_\Delta$ in \Cref{eq:pf_function_of_alpha} versus $\mathcal{P}_\mathrm{T}$ and $\mathcal{P}_\mathrm{{}^3He}$ (both are varied together) for each curve, $p_\mathrm{f}$ is normalized to the maximum power density (in $\mathcal{P}_\mathrm{T}$ and $\mathcal{P}_\mathrm{{}^3He}$) for $P_\mathrm{D}, P_\mathrm{{}^3He}, \alpha = 1, 1, 1/3$. (a) T = 20 keV, (b) T = 50 keV. The $\alpha$ values in second legend column correspond to the optimal $\alpha$ for maximizing $p_\Delta$ for $\mathcal{P}_\mathrm{T}=\mathcal{P}_\mathrm{{}^3He}= 1$.}
    \label{fig:pDelta_versus_inherited_spin}
\end{figure}

\begin{figure}[tb!]
    \centering
    \begin{subfigure}[t]{0.99\textwidth}
    \centering
    \includegraphics[width=1.0\textwidth]{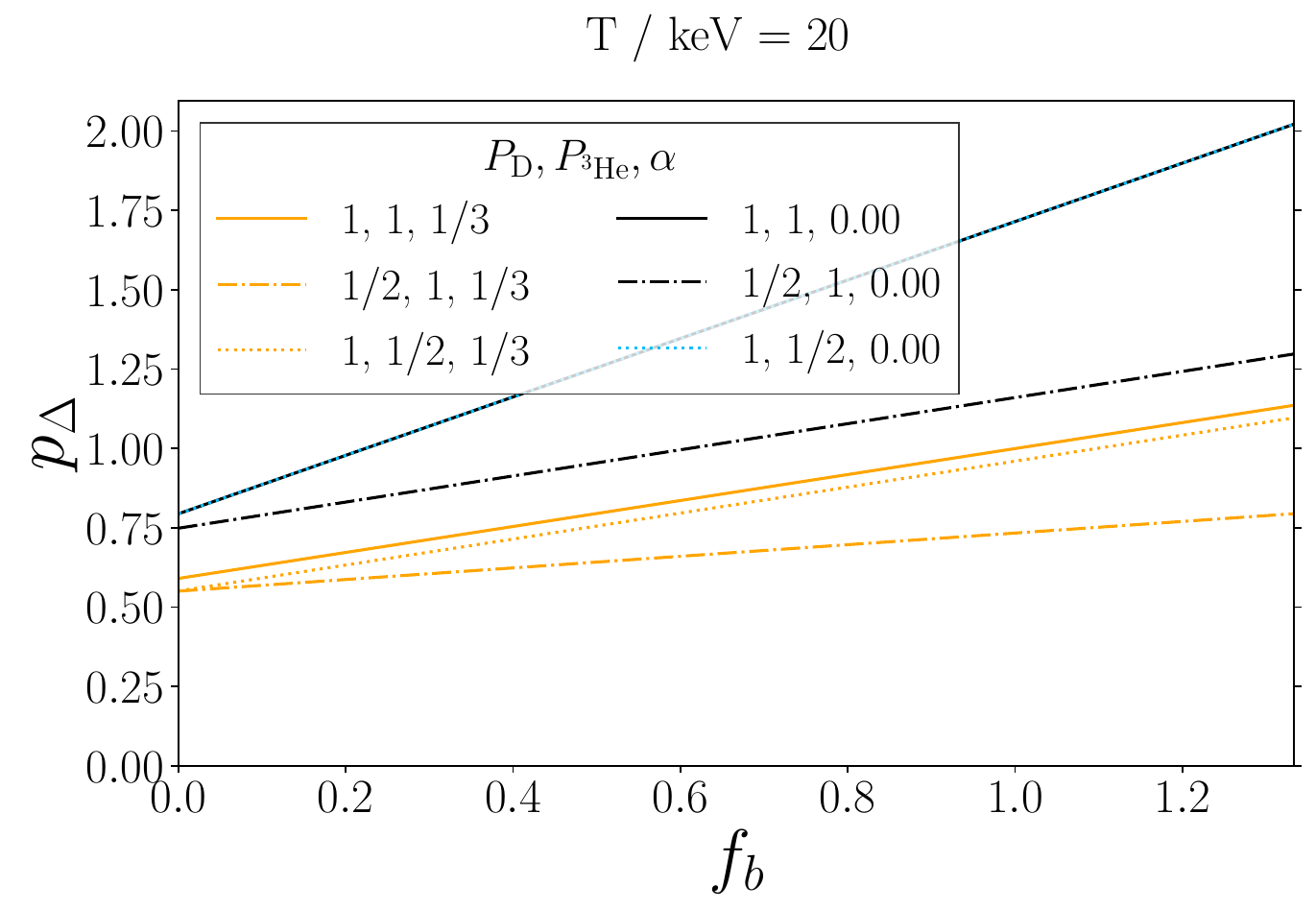}
    \caption{}
    \end{subfigure}
    \centering
    \begin{subfigure}[t]{0.99\textwidth}
    \centering
    \includegraphics[width=1.0\textwidth]{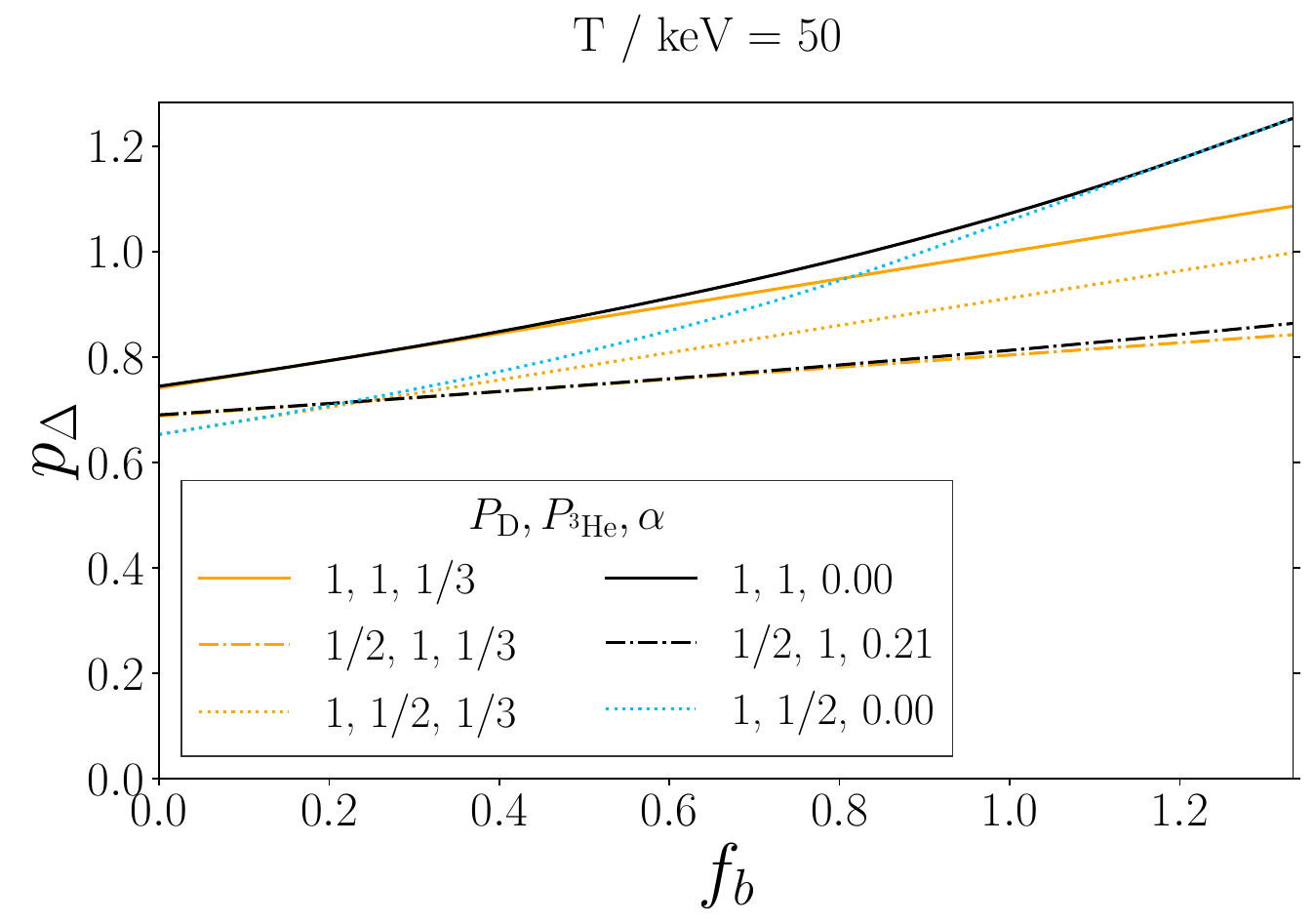}
    \caption{}
    \end{subfigure}
    \caption{Power density enhancement/suppression $p_\Delta$ in \Cref{eq:pf_function_of_alpha} versus D-D branching ratio $f_b$, $p_\mathrm{f}$ is normalized to the  power density for $P_\mathrm{D}, P_\mathrm{{}^3He}, \alpha, f_b = 1, 1, 1/3, 1$. (a) T = 20 keV, (b) T = 50 keV. The $\alpha$ values in second legend column correspond to the optimal $\alpha$ for maximizing $p_\Delta$ for the maximum $f_b$ value shown on the x-axis. We assumed a 100\% burn-up for secondary D-${}^3$He and tritium reactions. $f_b$ is varied at constant $\kappa_\mathrm{DD,p} = 0.5$ (see \Cref{eq:DD_branching}).}
    \label{fig:pDelta_versus_branching_ratio}
\end{figure}

\begin{figure*}[tb!]
    \centering
    \begin{subfigure}[t]{0.94\textwidth}
    \centering
    \includegraphics[width=1.0\textwidth]{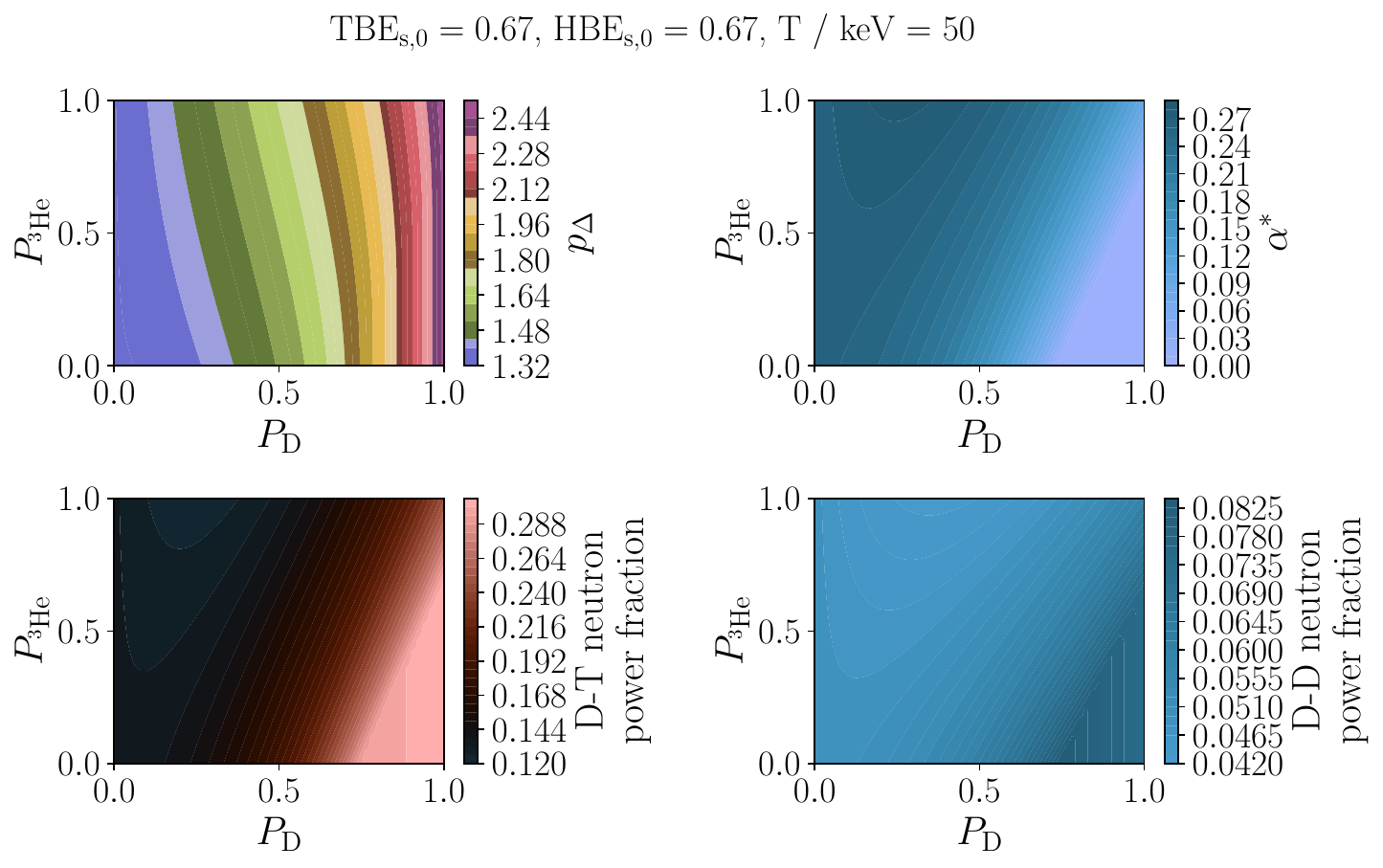}
    \end{subfigure}
    \caption{D-${}^{3}$He system with D and He3 fuel ratio optimized for maximum fusion power. Top left: Fusion power multiplier $p_\Delta$ (\Cref{eq:pDeltaform1}) for deuterium and helium-3 vector polarization $P_\mathrm{D}$ and $P_\mathrm{{}^{3}He}$, relative to power from unpolarized D-${}^{3}$He reactions. Top right: optimal $\alpha = 2n_\mathrm{{}^{3}He}/n_e$ value for maximizing the fusion power. Bottom left, bottom right: fraction of thermal fusion power coming from D-T and D-D neutrons. Here $\mathrm{TBE}_\mathrm{s,0} = \mathrm{HBE}_\mathrm{s,0} = 0.67$ and T/keV = 50.}
    \label{fig:temperaturedependent_optimized_alpha_branch_ratios}
\end{figure*}

In \Cref{fig:temperaturedependent_optimized_alpha}, we show $p_\Delta$, the optimal $\alpha$ value, $\alpha*$, and the corresponding power fractions for four temperatures with $\mathrm{TBE}_\mathrm{s} = \mathrm{HBE}_\mathrm{s} = 0.50$. We assume that only 50\% of the spin-polarization carries from D-D reactions to the resulting T and ${}^{3}$He, which we discuss in detail in the next section.

\subsection{Inherited Spin Polarization} \label{sec:inherited_spin}

A subtlety arises if the newly produced T or $^3$He inherits some polarization from its parent D-D reaction. For instance, D-D fusion reactions with polarized deuterons produce tritium, which itself might have some polarization. This polarized tritium can further enhance secondary channels like D-T or T-$^3$He. While the inherited spin is well-known for D-T and D-${}^3$He fusion products \cite{Kulsrud1982}, it is not well-characterized for D-D reactions.

In a simplified approach, we define a fraction $\mathcal{P}_\mathrm{T}$ as the effective polarization carried by T produced in the D-D (p) branch, and a fraction $\mathcal{P}_\mathrm{He3}$ for the $^3$He produced in the D-D (n) branch. If the newly created T or $^3$He from D-D is partially spin-polarized, we incorporate that by letting
\begin{equation}
P_\mathrm{T}= \mathcal{P}_\mathrm{T} P_\mathrm{D},\quad\quad P_\mathrm{He3,s} = \mathcal{P}_\mathrm{He3} P_\mathrm{D},
\label{eq:transfercoefficients}
\end{equation}
for some transfer coefficients $\mathcal{P}_\mathrm{T}$, $\mathcal{P}_\mathrm{He3}$. Then for the secondary channels D-T or D-$^3$He, the reactivity is
\begin{equation}
\begin{aligned}
&\langle \sigma v \rangle_{\mathrm{DT}} = \langle \overline{\sigma} v \rangle_{\mathrm{DT}} \bigl[1 + \frac{\mathcal{P}_\mathrm{T} P_\mathrm{D}^2}{2} \bigr], \\
&\langle \sigma v \rangle_{\mathrm{D{}^3He}} = \langle \overline{\sigma} v \rangle_{\mathrm{D{}^3He,s}} \bigl[1 + \frac{\mathcal{P}_\mathrm{{}^3He} P_\mathrm{D}^2}{2} \bigr]. \\
\end{aligned}
\end{equation}
Hence, a further $P_\mathrm{D}^2$ factor arises if T is created from polarized D–D. This can provide an additional multiplicative boost to the secondary channel, but also changes how the optimum fueling fraction $\alpha$ might shift. We also update the burn efficiency model in \Cref{eq:burn_efficiencies} to include effects of spin inheritance,
\begin{equation}
\begin{aligned}
    & \mathrm{HBE}_\mathrm{s} = \mathrm{HBE}_\mathrm{s,0} \left( 1 + 
    \frac{\mathcal{P}_\mathrm{{}^{3}He} P_\mathrm{D}^2}{2} \right) \\
    & \mathrm{TBE}_\mathrm{s} = \mathrm{TBE}_\mathrm{s,0} \left( 1 + \frac{\mathcal{P}_\mathrm{T} P_\mathrm{D}^2}{2} \right),
\end{aligned}
\label{eq:burn_efficiencies_inheritance}
\end{equation}
In \Cref{fig:pDelta_versus_inherited_spin}, we plot $p_\Delta$ for different values of $\mathcal{P}_\mathrm{T}$, $\mathcal{P}_\mathrm{He3}$ for polarized fuel, demonstrating the effect of the spin transfer on the power density. In these figures, the effect of spin transfer is captured through \Cref{eq:burn_efficiencies_inheritance}. In \Cref{fig:pDelta_versus_inherited_spin}(a), the fusion power is independent of $P_\mathrm{{}^3He}$ for the optimized $\alpha$ series (black and blue) because at T = 20 keV, the optimal $\alpha$ value is $\alpha = 0$. Therefore, all of the power comes from D-D reactions and the resulting secondary reactions. In \Cref{fig:pDelta_versus_inherited_spin}(b), we consider a higher temperature, T = 50 keV, where a significant fraction of the power comes from primary D-${}^3$He fusion reactions. Therefore, here the system is less sensitive to the  $\mathcal{P}_\mathrm{T}$, $\mathcal{P}_\mathrm{He3}$ transfer coefficients: for T = 20 keV, the plasma is D-D reaction-dominated and the optimized $\alpha$ fully polarized curve has a power density 70\% higher than the non-optimized $\alpha$ fully polarized curve. At T = 50 keV, there is only a modest 10\% power boost from optimizing $\alpha$ for the fully polarized branch. Finally, it is worth noting that because both D-D fusion branches produce one spin 1/2 and one spin 1 particle, we have used $\mathcal{P}_\mathrm{T}$ = $\mathcal{P}_\mathrm{He3}$ -- while this might be an incorrect assumption, at the time of writing this work we are unaware of measurements contradicting this.

\subsection{Deuterium Branching Ratios} \label{sec:branch_ratios}

In the above, we assumed that the reactivity for the two deuterium branches was equal. However, predictions for the effect of spin polarization on the D-D reactivity typically find different reactivities for the two D-D branches \cite{Engels_2014}. We now consider cases where there is an asymmetry between the reactivities for the two branches. In our model, the branching ratio of the two D-D reactions is
\begin{equation}
    f_b = \frac{\langle \overline{\sigma} v \rangle_{\mathrm{DD,n}}}{\langle \overline{\sigma} v \rangle_\mathrm{DD,p}} \frac{1 + \kappa_\mathrm{DD,n} P_\mathrm{D}^2}{1 + \kappa_\mathrm{DD,p} P_\mathrm{D}^2},
    \label{eq:DD_branching}
\end{equation}
which is equal to 1/2 for unpolarized D-D fusion. In \Cref{fig:pDelta_versus_branching_ratio} we plot the fusion power versus $f_b$ by varying $\kappa_\mathrm{DD,n}$ at fixed $\kappa_\mathrm{DD,p} = 0.5$. In plasmas where there is significant fusion power arising from D-D reactions, the branching ratio has a big impact.

\subsection{Optimized Fuel Ratio: Minimize D-T Neutrons} \label{sec:minimize_DT_neutrons}

There are significant engineering challenges arising from the 14 MeV neutrons born from D-T fusion reactions. In this section, we examine possible paths to minimizing the D-T neutrons in a D-${}^3$He plasma and how spin polarization can play a role.

There are at least three paths for minimizing D-T neutrons:
\begin{enumerate}
    \item Minimize the number of D-T reactions by reducing the D+D $\to$ p+T channel by optimizing temperature, D-${}^{3}$He fuel mix, and the deuterium polarization.
    \item Preferentially remove the tritium before it undergoes a fusion reaction \cite{Sheffield_2008}.
    \item In a spin-polarized D-${}^3$He plasma with $\mathcal{P}_\mathrm{D} = 1$, preferentially transfer spin to T such that is has $\mathcal{P}_\mathrm{D} = -1$ with a transfer coefficient $\mathcal{P}_\mathrm{T} = -1$ (see \Cref{eq:transfercoefficients}). This reduces the D-T fusion reactivity.
\end{enumerate}

\begin{figure*}[tb!]
    \centering
    \begin{subfigure}[t]{0.99\textwidth}
    \centering
    \includegraphics[width=1.0\textwidth]{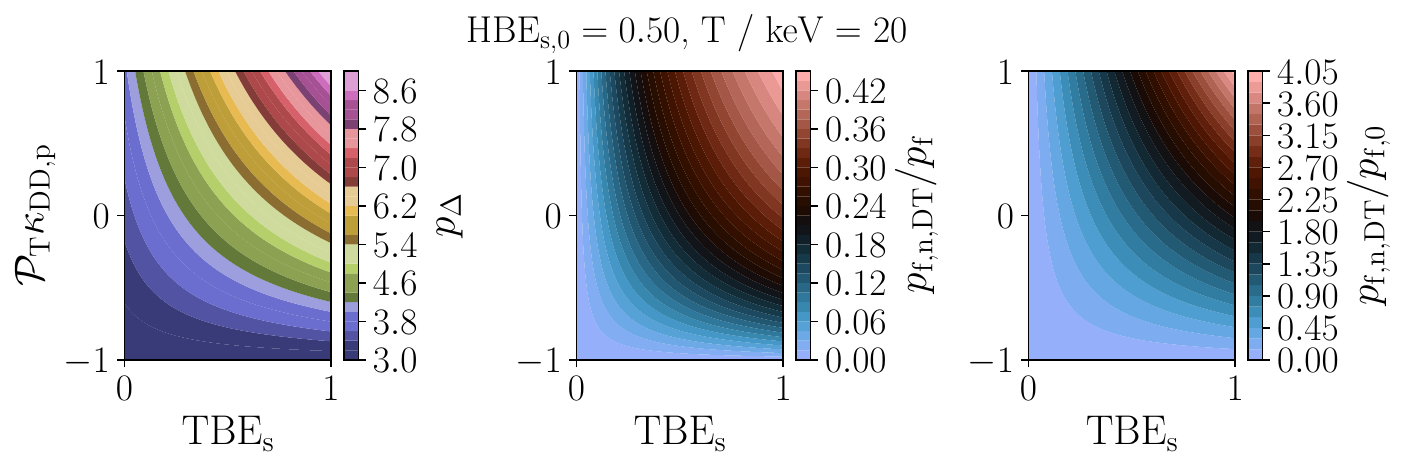}
    \end{subfigure}
    \caption{(a) Power density enhancement/suppression $p_\Delta$ in \Cref{eq:pf_function_of_alpha} versus $\mathrm{TBE_s}$ and $\mathcal{P}_\mathrm{T} \kappa_{DD,p}$, (b) D-T neutron power fraction of the total fusion power, (c) D-T neutron power fraction of the unpolarized D-${}^3$He fusion power. Here $P_\mathrm{D} P_\mathrm{{}^3He} = 1.$, $\mathrm{HBE_s} = 0.5$, and $\kappa_\mathrm{DD,n} = 0.5$.}
    \label{fig:pDelta_versus_TBEs_kappaDDp}
\end{figure*}

We focus on prospects for these three cases by varying $\mathcal{P}_\mathrm{T} \kappa_\mathrm{DD,p}$ and $\mathrm{TBE_s}$; in \Cref{fig:pDelta_versus_TBEs_kappaDDp}, we plot $p_\Delta$, the fraction of the fusion power from D-T neutrons $p_\mathrm{f,n,DT} / p_\mathrm{f}$, and the the fraction of the fusion power from D-T neutrons relative to the nominal power $p_\mathrm{f,n,DT} / p_\mathrm{f,0}$. In the `maximally neutronic' limit of $\mathcal{P}_\mathrm{T} \kappa_\mathrm{DD,p} \to 1$ and $\mathrm{TBE_s} \to 1$, $p_\mathrm{f,n,DT} / p_\mathrm{f}$ tends to roughly 40\%. The condition for zero D-T neutrons is $\mathcal{P}_\mathrm{T} \kappa_\mathrm{DD,p} = 0$ or $\mathrm{TBE_s} = 0$. From the perspective of maximizing fusion power, $\mathrm{TBE_s} = 0$ is preferable because the D-D $\to$ p + T reaction is permitted but all of the tritium escapes the plasma. However, from the perspective of minimizing tritium handling and inventory, $\mathcal{P}_\mathrm{T} \kappa_\mathrm{DD,p} = 0$ is preferable because no tritium is produced from  D-D $\to$ p + T reactions.

Next, we show how spin polarization can help achieve a more aneutronic burn. Achieving a pure aneutronic burn in D-${}^{3}$He plasmas might be desirable because it eliminates the need for any neutron shielding. However, a pure aneutronic burn is challenging because it generally requires suppressing both D-D fusion channels -- the D+D $\to$ p+T channel to prevent D-T fusion, and the D+D $\to$ n+He-3 channel to prevent 2.5 MeV neutrons. Fortunately, 2.5 MeV neutrons are easier to shield than 14.1 neutrons from D-T reactions because of extensive engineering experience with shielding 2.5 MeV neutrons. 

There are several strategies for obtaining aneutronic burn, all of them highly speculative. The first option is using spin polarization to suppress both D-D branches -- this is highly speculative because the effect of spin polarization on polarized D-D fusion is not yet known. The second option is using spin polarization to suppress the D+D $\to$ n+He-3 reaction and actively filtering out the tritium produced from the D+D $\to$ p+T reaction. This is challenging, potentially risky (D-T neutrons could be produced if the D+D $\to$ p+T reaction channel is not fully suppressed for whatever reason), and involves dealing with tritium. The third strategy is to maintain the helium-3 temperature much higher than the deuterium temperature. In addition to such a temperature difference being exceedingly challenging to maintain, it would also reduce the D-${}^{3}$He reactivity. This would put further downward pressure on the already low power density from D-${}^{3}$He fusion reactions.

\begin{figure*}[bt!]
    \centering
    \begin{subfigure}[t]{0.99\textwidth}
    \centering
    \includegraphics[width=1.0\textwidth]{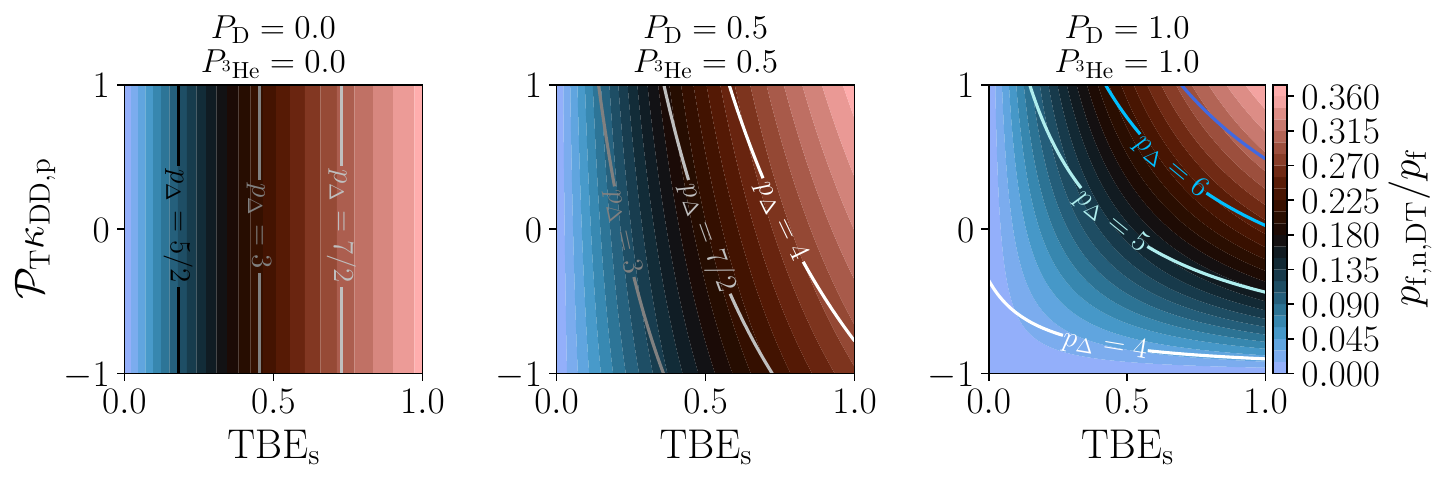}
    \end{subfigure}
    \caption{Power density enhancement/suppression $p_\Delta$ in \Cref{eq:pf_function_of_alpha} versus $\mathrm{TBE_s}$ and $\mathcal{P}_\mathrm{T} \kappa_{DD,p}$ for three combinations of $P_\mathrm{D} P_\mathrm{{}^3He} = 1$. Here $\mathrm{HBE_s} = 0.67$, $\kappa_\mathrm{DD,n} = 0.5$, and T = 20 keV.}
    \label{fig:pDelta_versus_TBEs_kappaDDp_diff_SPF_configs}
\end{figure*}

\section{Dual-Use Electric Power and Process Heat}

In this section, we describe the effect of spin-polarized fuel on the electric power and heat output of a hypothetical D-${}^3$He fusion power plant. Similar analyses have been performed for a D-T fusion power plant \cite{Heidbrink2024,Parisi_2025b}, showing how spin-polarized D-T could enable concepts that were previously on the margin of feasibility with unpolarized fuel to generate significant net electric power.

An FPP produces power from fusion reactions. The energy from fusion reactions is initially in the form of kinetic energy carried by neutrons and charged particles. The neutrons deposit their energy into a blanket and other structural material as heat that is converted into electricity with thermal conversion efficiency $\eta_{\text{th}}$. Charged particles have multiple pathways for depositing energy: they can deposit their energy in the plasma, in structural materials such as a first wall, or their energy can be directly converted to electricity via DEC with efficiency $\eta_\mathrm{DC}$.

Because the thermal conversion efficiency $\eta_{\text{th}}$ is less than 100\%, there is always some relatively low-temperature heat that is either exhausted to the environment or can be used for heating or low-temperature process heat. However, fusion may also produce high-temperature process heat that is valuable for energy-intensive manufacturing processes. Therefore there are plausible scenarios where FPPs produce both net electric power $P_\mathrm{net}$ and high-grade heat $Q_\mathrm{req}$.

DEC enables the transformation of kinetic energy from charged fusion products into electricity, bypassing traditional thermal cycles. Early implementations included MHD generators, which extract power from conducting fluids moving through magnetic fields \cite{rosa1987magnetohydrodynamic}. Modern DEC techniques are typically electrostatic or electromagnetic. Electrostatic systems such as traveling-wave and grid-based decelerators use electric fields to slow and collect fast ions, converting their energy into high-voltage DC \cite{tomita1993direct}. Electromagnetic approaches capture energy from time-varying magnetic fields generated by escaping particles or expanding plasmas \cite{wurden2016magneto}.

DEC is especially promising for advanced fuels like D-${}^{3}$He and p-${}^{11}$B, where most fusion energy is carried by charged particles rather than neutrons. In D-${}^{3}$He plasmas, for example, 14.7 MeV protons and 3.6 MeV alphas can be decelerated electrostatically for efficient energy recovery \cite{rostoker1997colliding}. In addition to higher electricity conversion efficiency, DEC is simpler than a steam cycle, potentially reducing the number of plant components and decreasing maintenance complexity. There are notable objections to DEC - for example, in tokamaks using D-${}^{3}$He fuel, the majority of fusion power is predicted to arrive at the first wall as heat via bremsstrahlung and synchrotron radiation, leaving little practical margin to extract significant power through direct conversion without imposing unrealistic plasma impurity or confinement conditions \cite{Stott_2005}.

We consider two FPP scenarios: a D-T plant and a D-${}^{3}$He plant. From the total fusion power $P_\mathrm{f}$ a certain amount must be recirculated to power the plant systems with electricity requirement $P_\mathrm{recirc}$. The FPP may feature DEC that allows the fraction $\gamma$ of the fusion power carried by charged particles to be converted directly to electricity; in a $T\sim$10-20 keV D-T plasma, $\gamma$ is very close to zero. The remaining fusion power is thermalized and converted to electricity with efficiency $\eta_{\text{th}}$.

\noindent
\textbf{D-T plant:} 
All fusion power is assumed to become heat in the blanket and surrounding structures, so the net electric power is given by
\begin{equation}
P_\mathrm{net}^{(\mathrm{D\text{-}T})}
= \eta_{\text{th}}\bigl(P_{\text{f}} - Q_{\text{req}}\bigr)
- P_\mathrm{recirc}
\end{equation}
where $P_\mathrm{f} - Q_{\text{req}}$ is the heat 
converted to electric power and the remaining $Q_{\text{req}}$ is delivered as heat. In order to simplify the calculation we omit the energy from neutron capture events in a blanket -- see \cite{Heidbrink2024,Parisi_2025b} for spin-polarized D-T fuel with a blanket. We assume that the plasma temperature is sufficiently low for the fusion power from other fusion reactions to be negligible.

\begin{figure*}[bt!]
    \centering
    \begin{subfigure}[t]{0.48\textwidth}
    \centering
    \includegraphics[width=1.0\textwidth]{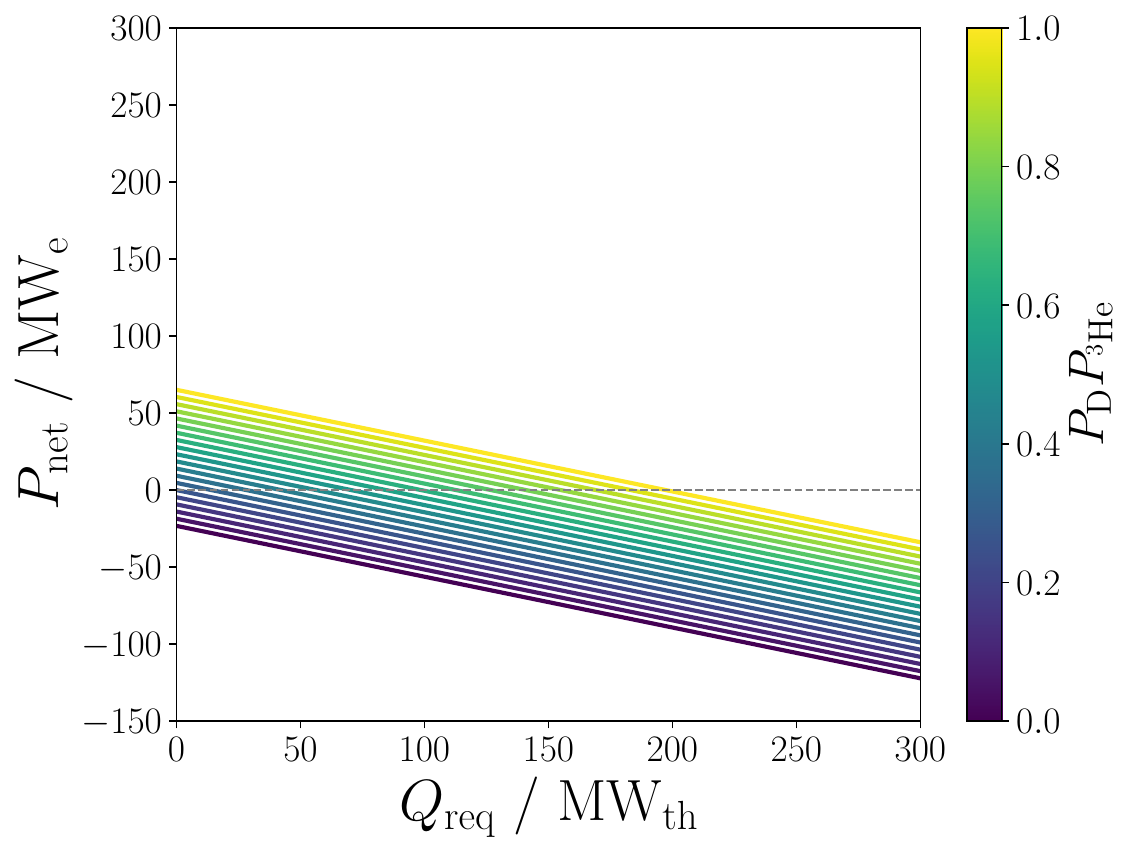}
    \caption{D-${}^{3}$He reactions only.}
    \end{subfigure}
    \centering
    \begin{subfigure}[t]{0.48\textwidth}
    \centering
    \includegraphics[width=1.0\textwidth]{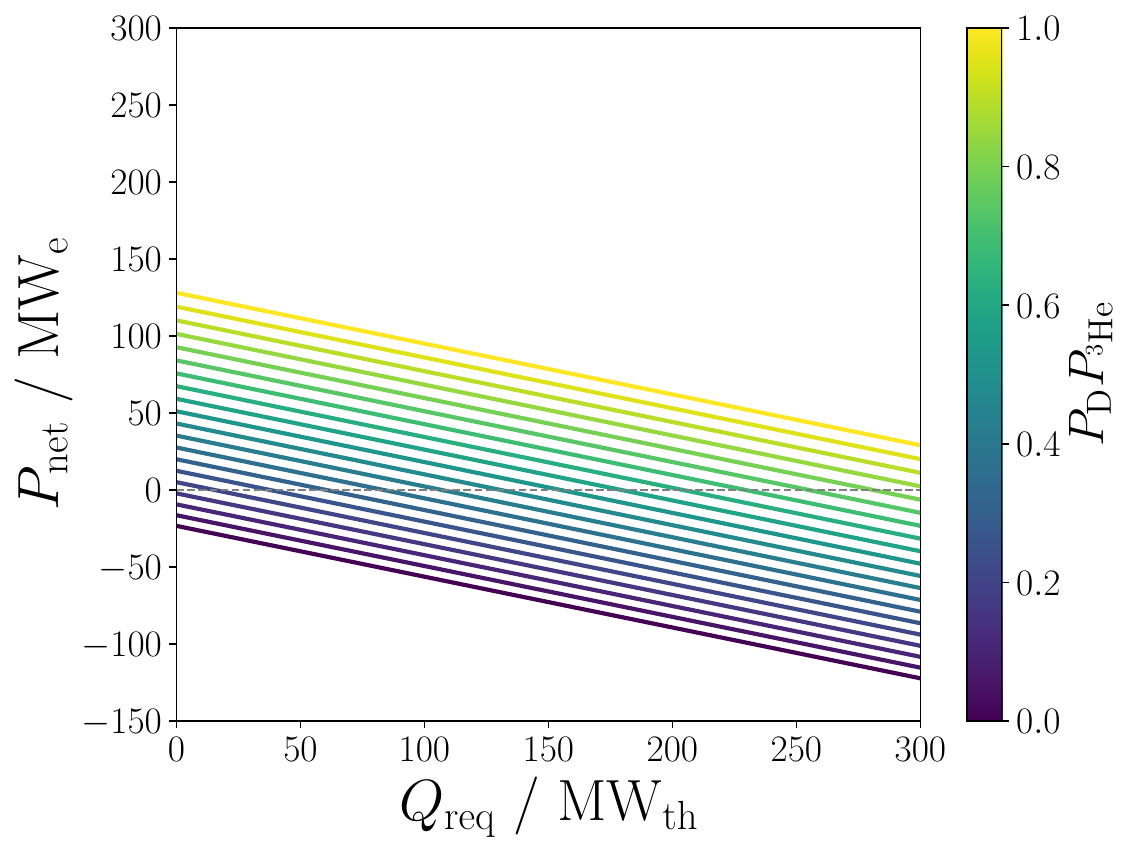}
    \caption{(a) + $\eta_\mathrm{DC} = 0.7 + 0.2 P_\mathrm{D} P_\mathrm{{}^{3}He}$.}
    \end{subfigure}
    \centering
    \begin{subfigure}[t]{0.48\textwidth}
    \centering
    \includegraphics[width=1.0\textwidth]{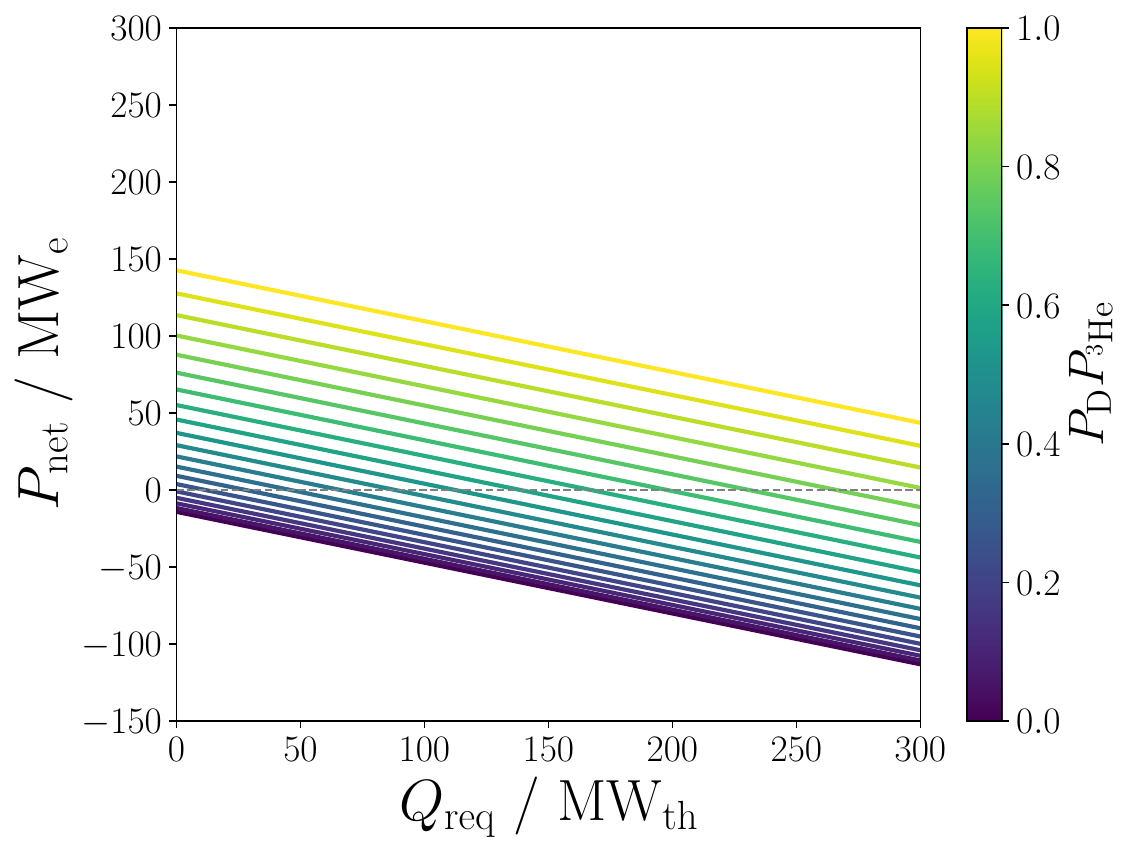}
    \caption{(b) + D-D fusion reactions.}
    \end{subfigure}
    \centering
    \begin{subfigure}[t]{0.48\textwidth}
    \centering
    \includegraphics[width=1.0\textwidth]{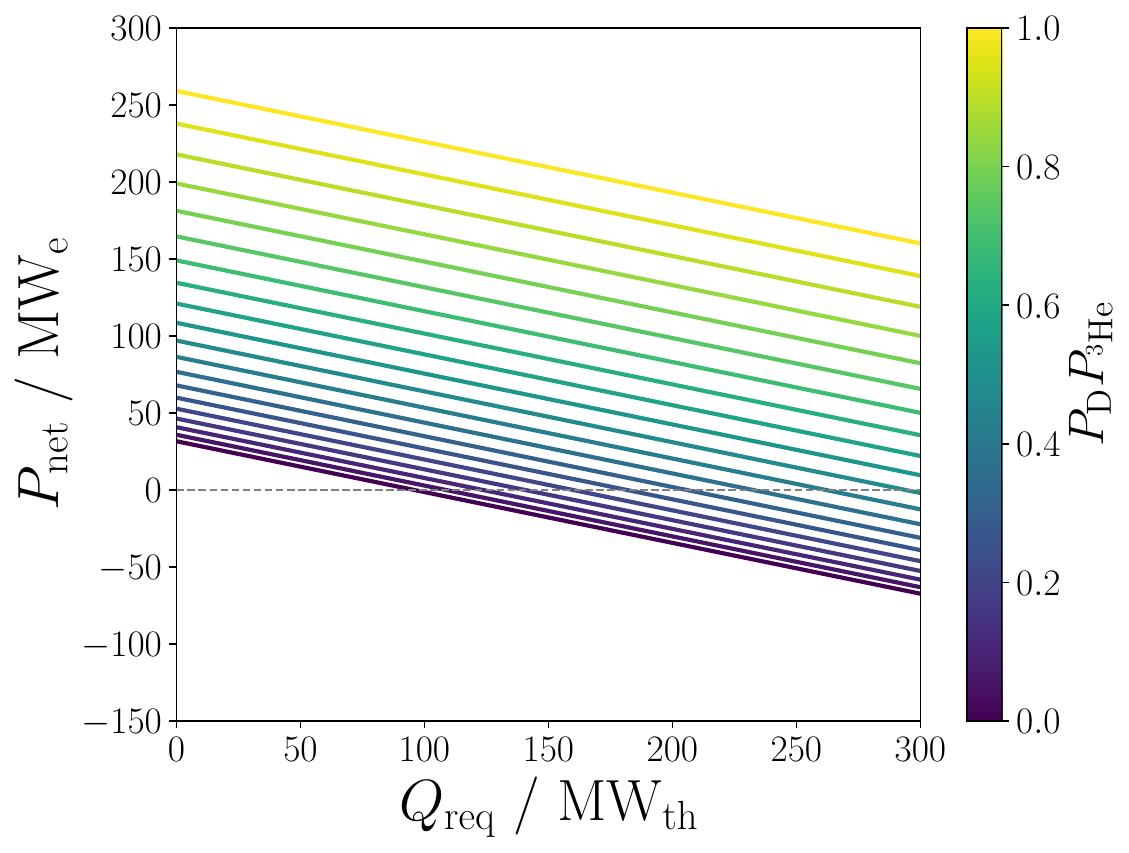}
    \caption{(c) + D-T and D-${}^{3}$He secondary reactions, optimized for $P_\mathrm{f}$.}
    \end{subfigure}
    \centering
    \begin{subfigure}[t]{0.48\textwidth}
    \centering
    \includegraphics[width=1.0\textwidth]{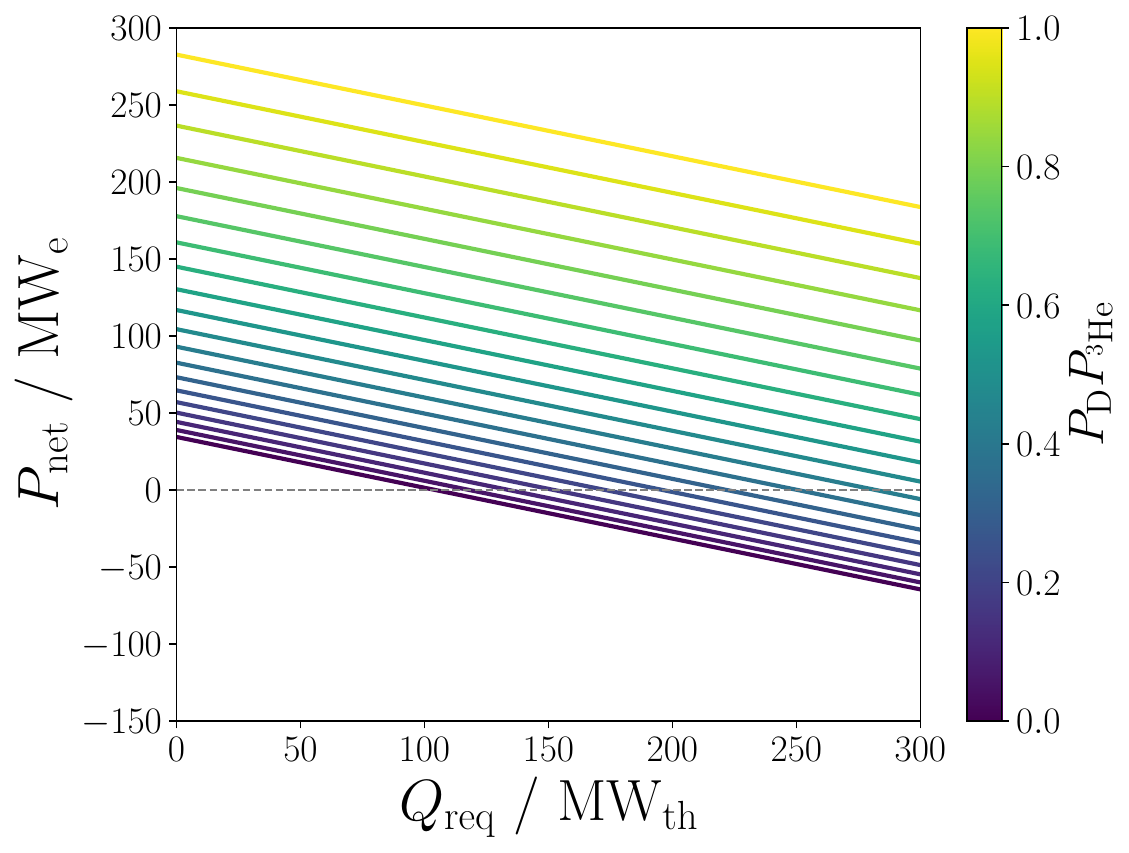}
    \caption{(c) + D-T and D-${}^{3}$He secondary reactions, optimized for $P_\mathrm{net}$.}
    \end{subfigure}
    \centering
    \begin{subfigure}[t]{0.48\textwidth}
    \centering
    \includegraphics[width=1.0\textwidth]{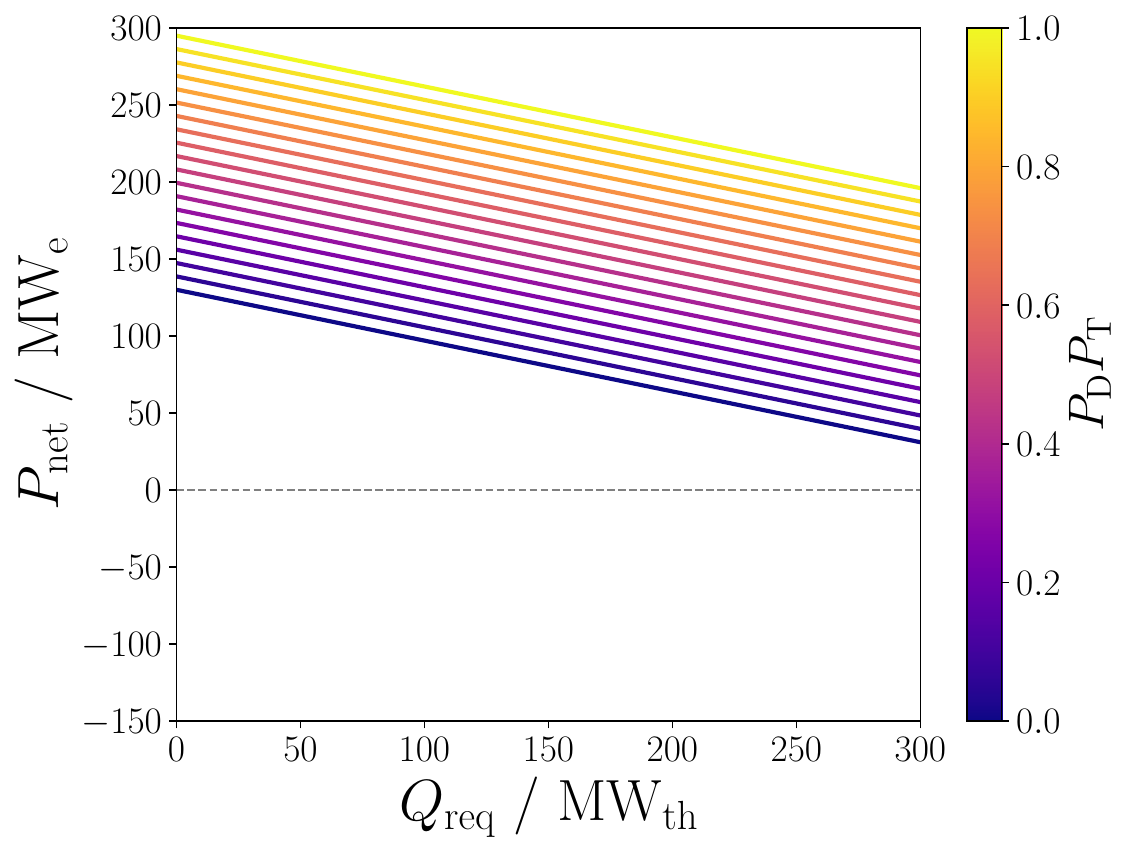}
    \caption{D-T reactions only, optimized for $P_\mathrm{net}$.}
    \end{subfigure}
    \caption{$P_{\text{net}}$ versus $Q_{\text{req}}$ for a D-${}^{3}$He plant with no secondary reactions with varying levels of additional effects. (a)
    For the base cases, we assume $\eta_{\text{th}} = 0.33$, $\eta_{\text{DC}} = 0.70$, $P_{\text{f}} = 300$ MW for D-${}^{3}$He and $P_{\text{f}} = 1000$ MW for D-T. Lines of different $P_\mathrm{D} P_ \mathrm{{}^{3}He}$ values are equally spaced in $P_\mathrm{D} P_ \mathrm{{}^{3}He}$. In both cases, $P_\mathrm{recirc} = 200$ MW.}
    \label{fig:Pnet_versus_Qnet_DHe3andDT}
\end{figure*}

\noindent
\textbf{D-${}^{3}$He plant:} 
In a D-${}^{3}$He plasma, a fraction $\gamma$ of all fusion power is carried by charged particles. We assume that a fraction $\gamma P_\mathrm{f}$ of the total fusion power is converted to electricity using direct energy conversion with efficiency $\eta_{\text{DC}}$. The remaining $\bigl(1 - \gamma\bigr) P_\mathrm{f}$ is deposited as heat. Therefore we require 
\begin{equation}
(1 - \gamma) P_\mathrm{f} \ge  Q_{\text{req}},
\end{equation}
to ensure there is enough thermal energy to meet the high-temperature process heat demand. 
Subtracting $Q_{\text{req}}$ from the thermal fraction, 
the leftover heat is converted at thermal efficiency $\eta_{\text{th}}$. In the D-${}^{3}$He plasma, we consider four reactions: D-${}^{3}$He, aneutronic and neutronic D-D, and D-T. We assume that D-${}^{3}$He direct converts $\gamma_\mathrm{D{}^{3}He}$ of its fusion power into electricity and the aneutronic D-D branch direct converts $\gamma_\mathrm{DD,p}$ of its fusion power into electricity. We assume the neutronic D-D branch and D-T deposit all power thermally. Thus, the net electric power from a D-${}^{3}$He plasma is
\begin{equation}
    \begin{aligned}
        P_\mathrm{net}^{(\mathrm{DHe^3})}
        = & \eta_{\mathrm{DC}} \gamma_\mathrm{D{}^{3}He} \left(P_{\mathrm{fus,D{}^{3}He}}+P_{\mathrm{fus,D{}^{3}He,s}}\right) \\
        + & \eta_{\mathrm{DC}} \gamma_\mathrm{DD,p} P_\mathrm{fus,DD,p} \\
        + & \eta_{\mathrm{th}}\Bigl[\bigl(1 - \gamma_\mathrm{D{}^{3}He} \bigr)\left(P_{\mathrm{fus,D{}^{3}He}}+P_{\mathrm{fus,D{}^{3}He,s}}\right) \\
        + & \bigl(1 - \gamma_\mathrm{DD,p}\bigr) P_\mathrm{fus,DD,p} + P_{\mathrm{fus,DD,n}} \\
        + & P_{\mathrm{fus,DT}} \Bigr] \\
        - & \eta_{\mathrm{th}} Q_{\mathrm{req}} - P_\mathrm{recirc}.
    \end{aligned}
    \label{eq:Pnet_DHe3}
\end{equation}
In our model, we choose the values $\gamma_\mathrm{D{}^{3}He} = 0.70$ and $\gamma_\mathrm{DD,p} = 0.33$ to account for some thermalization of charged particles occurring from the aneutronic D-${}^3$He and D-D reactions.

In \Cref{fig:Pnet_versus_Qnet_DHe3andDT}, we plot $P_{\text{net}}$ versus $Q_{\text{req}}$ for a D-${}^{3}$He plant (\Cref{fig:Pnet_versus_Qnet_DHe3andDT}(a)-(e)) and a D-T plant (\Cref{fig:Pnet_versus_Qnet_DHe3andDT}(f)). \Cref{fig:Pnet_versus_Qnet_DHe3andDT}(a)-(e) shows a progression of including increasingly more physical effects. \Cref{fig:Pnet_versus_Qnet_DHe3andDT}(a) shows the effect of vector polarization just on the D-${}^{3}$He net power and high-temperature process heat output. \Cref{fig:Pnet_versus_Qnet_DHe3andDT}(b) adds the effect of a variable direct energy conversion efficiency $\eta_\mathrm{DC} = 0.7 + 0.2 P_\mathrm{D} P_ \mathrm{{}^{3}He}$. \Cref{fig:Pnet_versus_Qnet_DHe3andDT}(c) adds D-D fusion reactions excluding secondary reactions. \Cref{fig:Pnet_versus_Qnet_DHe3andDT}(d) adds D-T and D-${}^{3}$He secondary reactions with a burn efficiency of $\mathrm{TBE}_\mathrm{s} = 0.67 + 0.33 P_\mathrm{D}^2$ and $\mathrm{HBE}_\mathrm{s} = 0.67 + 0.33 P_\mathrm{D}^2$, with the fuel ratio $\alpha$ optimized to maximize $P_\mathrm{f}$. \Cref{fig:Pnet_versus_Qnet_DHe3andDT}(e) is the same as (d) but the fuel ratio $\alpha$ is optimized to maximize the net power $P_\mathrm{net}$ rather than the total fusion power $P_\mathrm{f}$. 

Several important trends arise from this analysis. It is important to emphasize that we are presenting a highly simplified model of a FPP, and higher fidelity modeling is required for accurate results.

The first trend is the significant increase in net power generation capacity once D-D reactions are accounted for and once the fuel is polarized. Examination of \Cref{fig:Pnet_versus_Qnet_DHe3andDT}(a) shows a hypothetical D-${}^{3}$He plant only with D-${}^{3}$He fusion reactions, unpolarized fuel, and high-temperature process heat has a net electric power of $P_\mathrm{net} = - 25 $ MW, meaning it doesn't generate enough electricity to even power the plant's subsystems. However, when all of the effects of secondary D-D reactions are included, the same fuel is predicted to give a net electric power of $P_\mathrm{net} = 35 $ MW (\Cref{fig:Pnet_versus_Qnet_DHe3andDT}(e)), an increase of 60 MW. The effect with spin-polarized fuel is even starker: with fully polarized fuel, \Cref{fig:Pnet_versus_Qnet_DHe3andDT}(a) predicts that only including D-${}^{3}$He fusion reactions with $Q_\mathrm{req} = 0$ gives $P_\mathrm{net} = 65 $ MW. When all of the effects of secondary D-D reactions are included, the same fuel is predicted to give a net electric power of $P_\mathrm{net} = 285 $ MW (\Cref{fig:Pnet_versus_Qnet_DHe3andDT}(e)), an increase of 220 MW.

The second trend is that the marginal net electric power increase is nonlinear in the polarization -- this is seen by comparing the spacing of lines of constant $P_\mathrm{D} P_\mathrm{{}^3He}$ in \Cref{fig:Pnet_versus_Qnet_DHe3andDT}(b) and (c), where adding the D-D reactions in (c) from (b) already introduces a nonlinear dependence of $P_\mathrm{net}$ on $P_\mathrm{D} P_\mathrm{{}^3He}$. In contrast, examining the net power from a D-T plasma in \Cref{fig:Pnet_versus_Qnet_DHe3andDT}(f), the marginal increase in $P_\mathrm{net}$ is linear with $P_\mathrm{D} P_\mathrm{T}$ -- there are likely additional nonlinear effects of polarization that we have neglected in our analysis such as increased alpha particle heating that increases the plasma temperature and therefore modifies the reactivity \cite{Smith2018IAEA}. These effects would further increase the nonlinearity of $P_\mathrm{net}$ with $P_\mathrm{D} P_\mathrm{{}^3He}$ for D-${}^{3}$He plasmas, and introduce nonlinearity of $P_\mathrm{net}$ with $P_\mathrm{D} P_\mathrm{T}$ for D-T plasmas.

The third trend is that comparing the D-${}^{3}$He and D-T plasmas in \Cref{fig:Pnet_versus_Qnet_DHe3andDT}(e) and (f), in our model the polarization has a much bigger effect on $P_\mathrm{net}$ for D-${}^{3}$He than D-T plasmas -- comparing the $Q_\mathrm{req} = 0$ points for $P_\mathrm{D} P_\mathrm{{}^3He}$ and $P_\mathrm{D} P_\mathrm{T}$ we see that the net power increases by a factor of eight for the D-${}^{3}$He plasma but only a factor of two for the D-T plasma.

Another important feature is that in our model, $d P_\mathrm{net} / d Q_\mathrm{req}$ is always linear with value $- \eta_\mathrm{DC}$. This assumes that charged particles convert all of their energy directly into electricity -- we have not included the effects of allowing even more heat to be generated by thermalizing charged particles that were previously generating electricity with DEC. This may present another option for further increasing the output of high-temperature process heat $Q_\mathrm{req}$, but is not further explored here. 

Finally, we chose values of $\gamma_\mathrm{D{}^{3}He} = 0.70$ and $\gamma_\mathrm{DD,p} = 0.33$ to account for some thermalization of high energy charged particles. Increases in these values could further enhance $P_\mathrm{net}$.

\subsection{Engineering Gain with SPF} \label{sec:reactivity_energyrecovery}

\begin{figure*}[tb!]
    \centering
    \begin{subfigure}[t]{0.77\textwidth}
    \centering
    \includegraphics[width=1.0\textwidth]{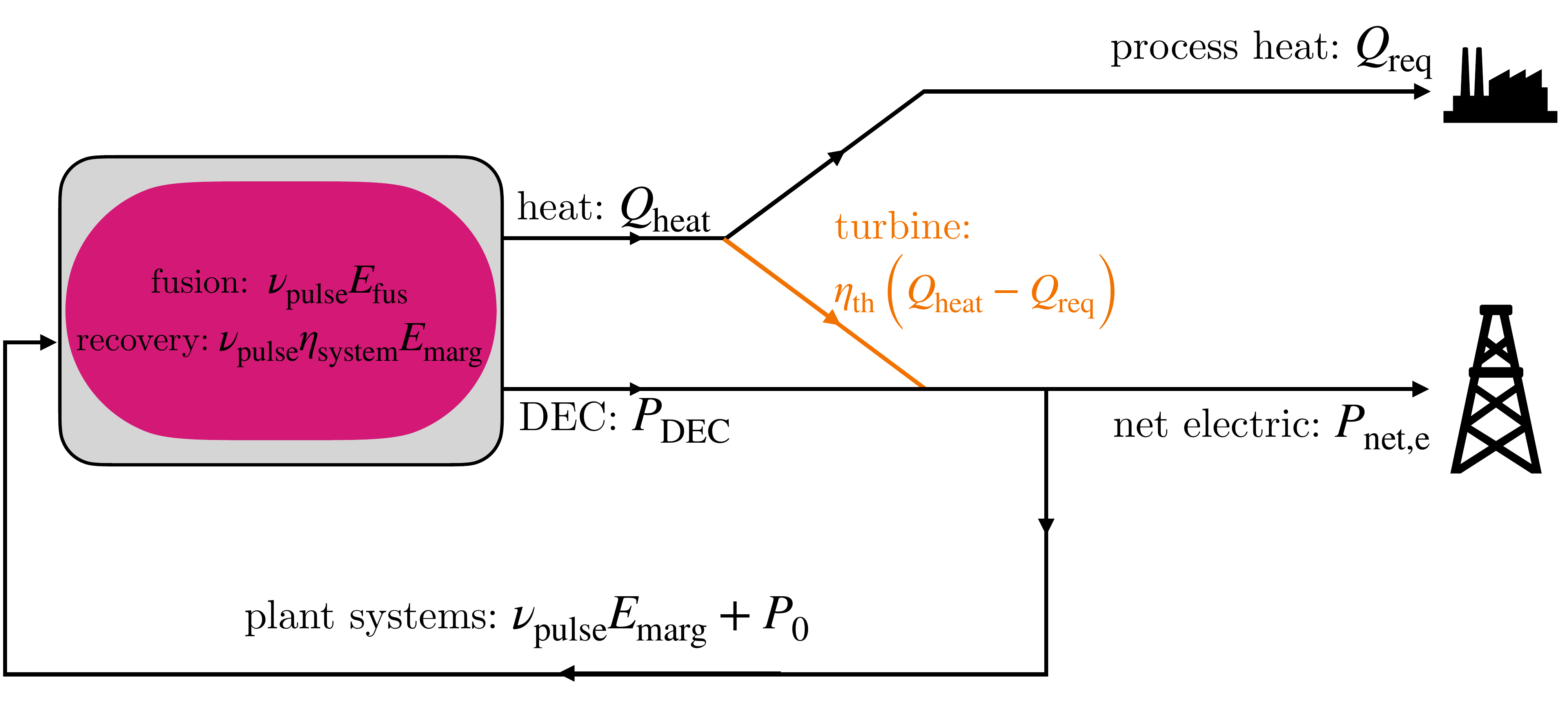}
    \end{subfigure}
    \caption{Schematic diagram for a pulsed power plant with direct energy conversion, energy recovery, and both electric and high-grade process heat output.}
    \label{fig:pulsed_plant_system}
\end{figure*}

In this section we show how SPF can have a large impact on engineering gain in a pulsed D-${}^{3}$He fusion device \cite{Kirtley_2023} -- these results can also be applied to a steady state device in a straightforward mathematical limit. 

Pulsed systems might be appealing because of the possibility to recover some power from the marginal energy per pulse \cite{Oliphant_1973}. DEC systems may also be easier to implement in pulsed systems because some DEC systems achieve higher efficiency in vacuum -- a pulsed system that largely evacuates the chamber per pulse might achieve higher DEC efficiency. The schematic for this system is shown in \Cref{fig:pulsed_plant_system}.

In this subsection, we neglect all fusion reactions except for the D-${}^{3}$He fusion reaction we assume that zero high-temperature process heat is required $E_\mathrm{req} = 0$. In the next subsection we include the effects of D-D and secondary reactions and use $E_\mathrm{req} > 0$.

We assume that fusion reactions occur in a pulsed manner with a frequency $\nu_\mathrm{pulse}$. The total energy released from fusion reactions in a single pulse is
\begin{equation}
E_\mathrm{f} = E_\mathrm{D{}^{3}He} N_\mathrm{{}^{3}He}^\mathrm{burn},
\end{equation}
where $N_\mathrm{{}^{3}He}^\mathrm{burn}$ is the number of helium-3 particles burned in a pulse. Over timescales much longer than the pulse length, the effective fusion power is therefore
\begin{equation}
P_\mathrm{f} = \nu_\mathrm{pulse} E_\mathrm{f} = \dot{N}_\mathrm{{}^{3}He}^\mathrm{burn} E_\mathrm{D{}^{3}He},
\end{equation}
where the helium-3 burn rate is
\begin{equation}
\dot{N}_\mathrm{{}^{3}He}^\mathrm{burn} \equiv \nu_\mathrm{pulse} N_\mathrm{{}^{3}He}^\mathrm{burn}. 
\end{equation}
The marginal energy cost to produce a pulse is $E_\mathrm{marg}$ and the background power cost (the continuous cost of running the machine without pulsing that includes cryoplant, basic systems, etc) is $P_0$. Therefore, the electricity cost of running the plant is
\begin{equation}
P_\mathrm{cost} = \nu_\mathrm{pulse} E_\mathrm{marg} + P_0. 
\end{equation}
The power from DEC conversion of the fusion products is
\begin{equation}
    P_\mathrm{DEC} = \eta_\mathrm{DC} E_\mathrm{f}.
\end{equation}
One proposed feature of pulsed schemes is their ability to recover a fraction $\eta_\mathrm{rec}$ of $E_\mathrm{marg}$ from each pulse. Over timescales longer than the pulse frequency, the recovered power is
\begin{equation}
    P_\mathrm{rec} = \eta_\mathrm{rec} E_\mathrm{marg}.
\end{equation}

We assume that both the proton and helium-4 from D-${}^3$He reactions are direct converted with equal efficiency. The claim of this section is that both $\eta_\mathrm{DC}$ and $E_\mathrm{f}$ can be modified using spin-polarized fuel relative to their nominal values $\eta_\mathrm{DC,0}$ and $E_\mathrm{f,0}$ for unpolarized fuel,
\begin{equation}
\eta_\mathrm{DC} = \mathcal{H} \eta_\mathrm{DC,0}, \;\;\; E_\mathrm{f} = \mathcal{N} E_\mathrm{f,0},
\label{eq:SPF}
\end{equation}
where $\mathcal{H}$ and $\mathcal{N}$ are functions depending on the polarization scheme and polarization fraction. For simplicity, we include all of the loss terms in $\eta_\mathrm{rec}$: Bremsstrahlung losses, cyclotron losses, transport losses, etc. These system losses may also be modified by polarization, $\eta_\mathrm{rec} = \mathcal{S} \eta_\mathrm{rec,0}$, where $\mathcal{S}$ depends on the polarization scheme and polarization fraction, but for simplicity in this work we use $\mathcal{S} = 1$. 

Combining these effects, the net electric power output is
\begin{equation}
P_{\mathrm{net} } = \nu_\mathrm{pulse} \left( E_\mathrm{marg} \left( \eta_\mathrm{rec} - 1 \right)  + \eta_\mathrm{DC} E_\mathrm{f} \right) - P_0.
\label{eq:Pnete}
\end{equation}

\begin{figure}[tb!]
    \centering
    \begin{subfigure}[t]{0.97\textwidth}
    \centering
    \includegraphics[width=1.0\textwidth]{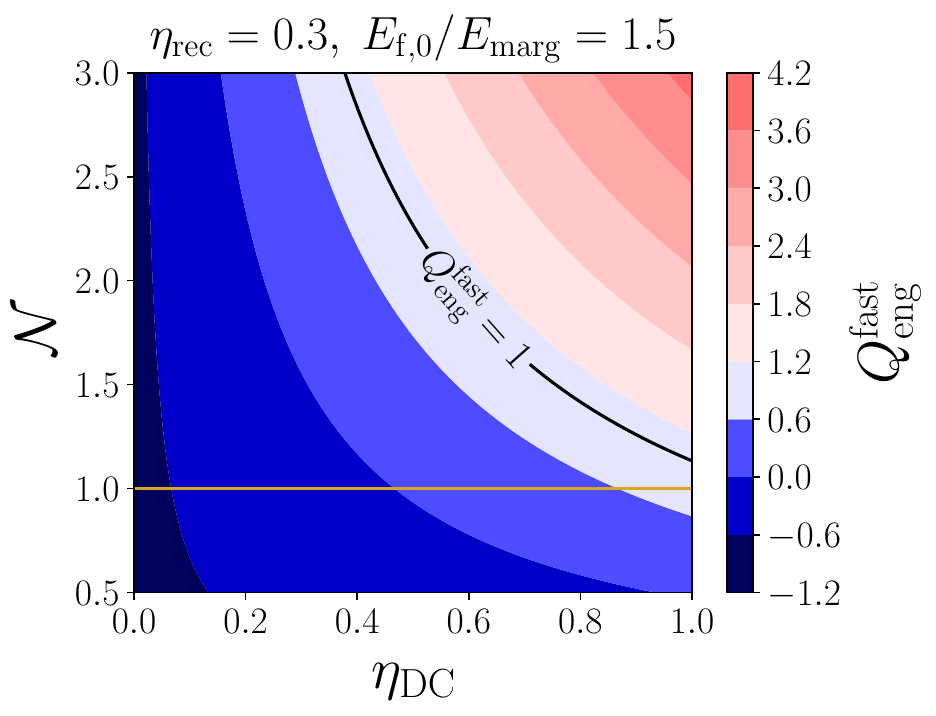}
    \caption{}
    \end{subfigure}
    \centering
    \begin{subfigure}[t]{0.97\textwidth}
    \centering
    \includegraphics[width=1.0\textwidth]{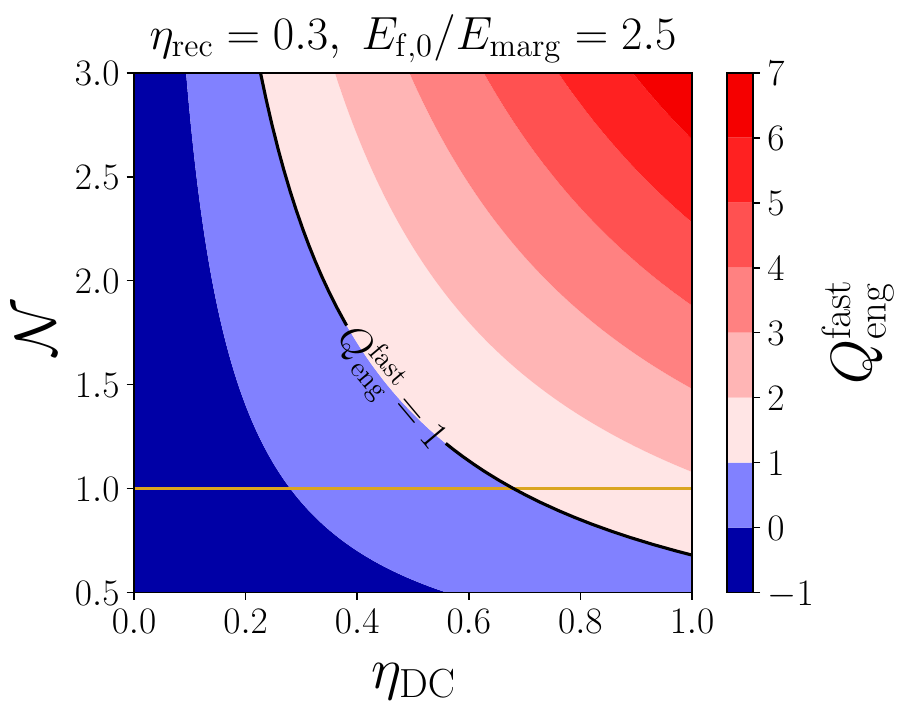}
    \caption{}
    \end{subfigure}
    \centering
    \begin{subfigure}[t]{0.97\textwidth}
    \centering
    \includegraphics[width=1.0\textwidth]{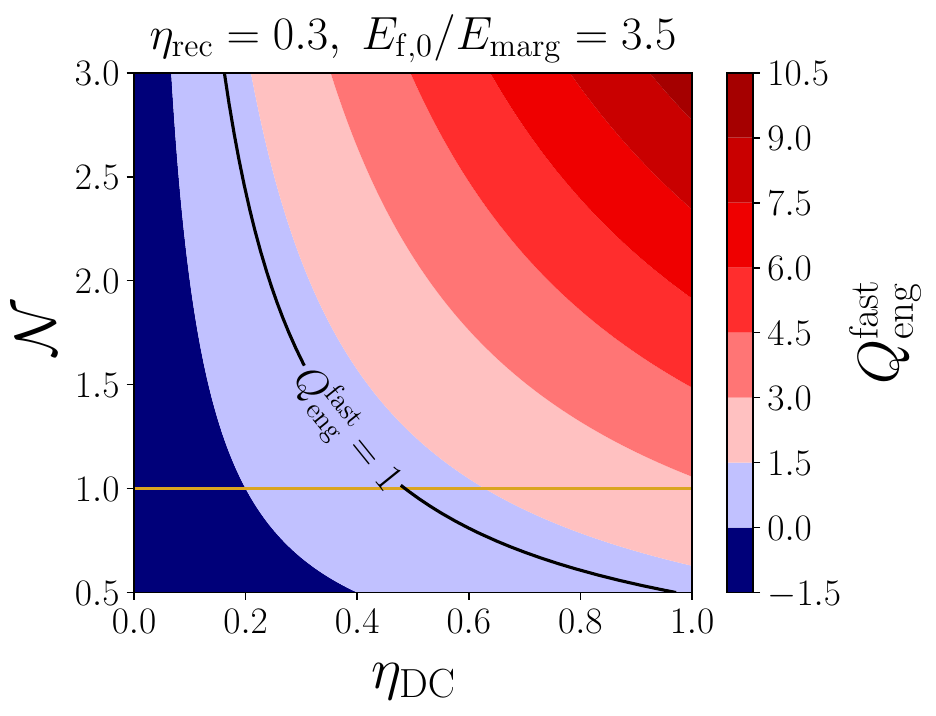}
    \caption{}
    \end{subfigure}
    \caption{$Q_\mathrm{eng}^\mathrm{fast}$ versus $\eta_\mathrm{DC}$ and $\mathcal{N}$ (see \Cref{eq:SPF}) for $\eta_\mathrm{rec} =0.3$. For (a), (b), (c), $E_\mathrm{f,0}/E_\mathrm{marg} = 1.5, 2.5, 3.5$. Horizontal golden line represents unpolarized fuel.}
    \label{fig:Qengineering_fast}
\end{figure}

\noindent
Assuming that the pulse frequency can be varied independently of other parameters, power breakeven, $P_\mathrm{net} \ge 0 $, can be specified in terms of a minimum pulse frequency $\nu_\mathrm{break}$,
\begin{equation}
\nu_\mathrm{pulse} > \nu_\mathrm{break} \equiv \frac{P_0}{E_\mathrm{marg} \left( \eta_\mathrm{rec} - 1 \right)  + \eta_\mathrm{DC} E_\mathrm{f}}. 
\end{equation}
We define engineering gain as
\begin{equation}
Q_\mathrm{eng} \equiv \frac{P_{\mathrm{net} }}{P_\mathrm{cost} }.
\end{equation}
Substituting \Cref{eq:Pnete} gives
\begin{equation}
Q_\mathrm{eng} = \frac{  \eta_\mathrm{rec} E_\mathrm{marg} + \eta_\mathrm{DC} E_\mathrm{f}}{P_0/\nu_\mathrm{pulse} + E_\mathrm{marg}} - 1.
\end{equation}

In the fast-pulse limit where the marginal energy cost per pulse greatly exceeds the background power $E_\mathrm{marg} \gg P_0/\nu_\mathrm{pulse}$, the engineering gain simplifies to
\begin{equation}
Q_\mathrm{eng}^\mathrm{fast} = \eta_\mathrm{rec} - 1 + \eta_\mathrm{DC} \frac{ E_\mathrm{f}}{E_\mathrm{marg} }.
\end{equation}
Substituting $E_\mathrm{f}$ from \Cref{eq:SPF} gives
\begin{equation}
Q_\mathrm{eng}^\mathrm{fast} = \eta_\mathrm{rec} - 1 + \eta_\mathrm{DC} \mathcal{N} \frac{ E_\mathrm{f,0}}{E_\mathrm{marg} }.
\end{equation}
We plot $Q_\mathrm{eng}^\mathrm{fast}$ versus $\mathcal{N}$ and $\eta_\mathrm{DC}$ in \Cref{fig:Qengineering_fast} for a fairly conservative set of parameters, $\eta_\mathrm{rec} =0.3$ and three values of $E_\mathrm{f,0}/E_\mathrm{marg}$: 1.5, 2.5, and 3.5. Increasing the fusion cross section with $\mathcal{N}$ or the efficiency of energy recovery with $\eta_\mathrm{DC}$ has a significant effect on the engineering gain. For all three values of $E_\mathrm{f,0}/E_\mathrm{marg}$, using spin-polarized fuel to increase $\mathcal{N}$ and $\eta_\mathrm{DC}$ makes it significantly easier to achieve engineering breakeven.

\subsection{Including Side and Secondary Reactions}

In this section, we generalize the pulsed scheme above to include side D-D reactions and secondary D-T and D-${}^{3}$He reactions. We also include the option of producing high-temperature process heat.

The total energy released per pulse is
\begin{equation}
    \begin{aligned}
    E_\mathrm{f} = & E_\mathrm{D{}^{3}He} N_\mathrm{{}^{3}He}^\mathrm{burn} + E_\mathrm{DD,p} \frac{N_\mathrm{D}^\mathrm{burn}}{4} + E_\mathrm{DD,n} \frac{N_\mathrm{D}^\mathrm{burn}}{4} \\
    + & E_\mathrm{DT} N_\mathrm{T}^\mathrm{burn} + E_\mathrm{D{}^{3}He} N_\mathrm{{}^3He,s},
    \end{aligned}
\end{equation}
where $N_\mathrm{D}^\mathrm{burn}$ is the number of deuterium nuclei burned in D-D reactions, $N_\mathrm{T}^\mathrm{burn}$ is the number of tritium nuclei burned in D-T reactions, and $N_\mathrm{{}^3He,s}$ is the number of helium-3 nuclei burned in secondary D-${}^{3}$He reactions. We assume that both branches of D-D reactions are burned in equal quantities.

\begin{figure}[bt!]
    \centering
    \begin{subfigure}[t]{0.98\textwidth}
    \centering
    \includegraphics[width=1.0\textwidth]{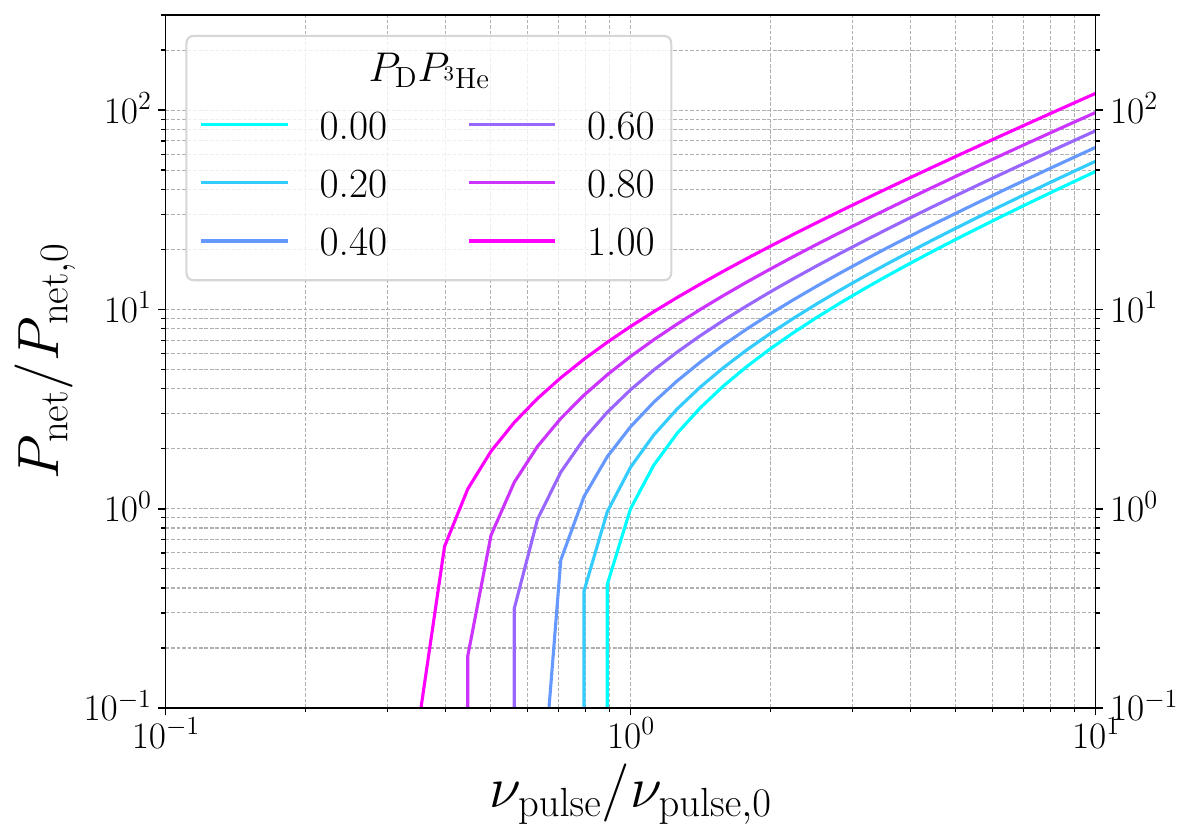}
    \caption{}
    \end{subfigure}
    \centering
    \begin{subfigure}[t]{0.98\textwidth}
    \centering
    \includegraphics[width=1.0\textwidth]{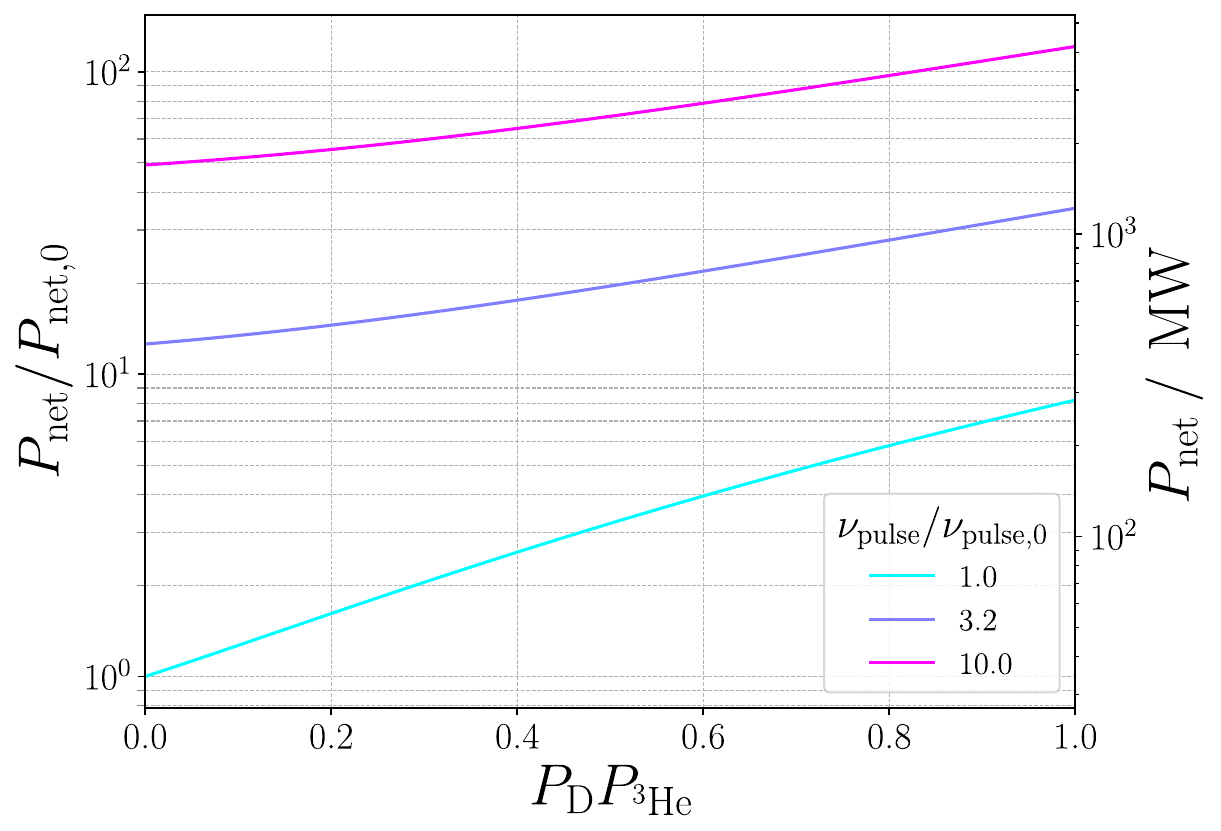}
    \caption{}
    \end{subfigure}
    \caption{Net electric power for a 50 keV D-${}^{3}$He plasma versus (a) $\nu_\mathrm{pulse}$ and (b) $P_\mathrm{D} P_\mathrm{{}^3He}$.}
    \label{fig:electric_fusion_gain_summary}
\end{figure}

\noindent
The effective fusion power is therefore
\begin{equation}
    \begin{aligned}        
    P_\mathrm{f} = & E_\mathrm{D{}^{3}He} \dot{N}_\mathrm{{}^{3}He}^\mathrm{burn} \\
    + & E_\mathrm{DD,p} \frac{ \dot{N}_\mathrm{DD}^\mathrm{burn}}{4} + E_\mathrm{DD,n} \frac{ \dot{N}_\mathrm{DD}^\mathrm{burn}}{4} \\
    + & E_\mathrm{DT} \dot{N}_\mathrm{T}^\mathrm{burn} + E_\mathrm{D{}^{3}He} \dot{N}_\mathrm{{}^3He,s}^\mathrm{burn}.
    \end{aligned}
    \label{eq:Pfus}
\end{equation}
The amount of energy recovered per pulse in direct conversion electricity is
\begin{equation}
    \begin{aligned}
        E_\mathrm{rec,e} = & \eta_\mathrm{rec} E_\mathrm{marg} \\
        + & \eta_{\mathrm{DC}} \gamma_\mathrm{D{}^{3}He} \left(E_{\mathrm{f,D{}^{3}He}} + E_\mathrm{f,D{}^{3}He,s}\right) \\
        + & \eta_{\mathrm{DC}} \gamma_\mathrm{DD,p} E_{\mathrm{fus,DD,p}},
    \end{aligned}
    \label{eq:Erecover_elec}
\end{equation}
and the amount recovered per pulse in electricity from heat after a required amount of heat $E_{\text{req}}$ is
\begin{equation}
    \begin{aligned}
        E_\mathrm{rec,Q} = & \eta_{\mathrm{th}}\Bigl[\bigl(1 - \gamma_\mathrm{D{}^{3}He} \bigr)\left(E_{\mathrm{f,D{}^{3}He}}+E_\mathrm{f,D{}^{3}He,s}\right) \\
        + & \bigl(1 - \gamma_\mathrm{DD,p}\bigr) E_{\text{fus,DD,n}} + E_\mathrm{f,DT} \Bigr] \\
        - & \eta_{\mathrm{th}} E_{\mathrm{req}}.
    \end{aligned}
    \label{eq:Erecover_heat}
\end{equation}
Therefore, over timescales longer than the pulse frequency the net electric power is
\begin{equation}
P_\mathrm{net} = \nu_\mathrm{pulse} \left( E_\mathrm{rec,e} + E_\mathrm{rec,Q} - E_\mathrm{marg} \right) - P_0,
\label{eq:netpower}
\end{equation}
and the high-temperature process heat output is $Q_\mathrm{req} = \nu_\mathrm{pulse} E_\mathrm{req}$.

In \Cref{fig:electric_fusion_gain_summary}(a), we plot the scaling of the net electric power gain $P_\mathrm{net} / P_\mathrm{net,0}$ versus $\nu_\mathrm{pulse} / \nu_\mathrm{pulse,0}$ for a pulsed device with a nominal pulse frequency of $\nu_\mathrm{pulse,0}$, $E_\mathrm{marg} = 5$ MJ, $P_\mathrm{0} = 150$ MW, $P_\mathrm{net,0} = 34$ MW, and no polarization $P_\mathrm{D} P_\mathrm{{}^3He}$. The fuel ratio $\alpha$ has been optimized to maximize $P_\mathrm{net}$. \Cref{fig:electric_fusion_gain_summary}(a) shows that fully polarized gives an increase in $P_\mathrm{net}$ of 5-10 compared with unpolarized fuel. Increasing the pulse frequency also has a significant nonlinear impact on $P_\mathrm{net} / P_\mathrm{net,0}$. In \Cref{fig:electric_fusion_gain_summary}(b) and (c) we plot the net electric versus polarization $P_\mathrm{D} P_\mathrm{{}^3He}$ for several $\nu_\mathrm{pulse} / \nu_\mathrm{pulse,0}$ values for the same pulsed device. Increasing the pulse frequency by a factor of ten and using fully spin-polarized D-${}^3$He increases the net predicted power by over a factor of 100 relative to the nominal power.

\section{Discussion} \label{sec:discussion}

In this work, we have calculated the effect of spin-polarized fuel in D-${}^3$He plasmas, accounting for D-${}^3$He, D-D, and secondary D-${}^3$He and D-T fusion reactions. We have demonstrated that under highly optimistic scenarios, spin-polarized fuel could increase the total fusion power by roughly an order of magnitude relative to the fusion power produced by unpolarized D-${}^3$He fusion reactions alone. Depending on the specifics of the D-${}^3$He fusion power plant, the net electric power can also increase by an order of magnitude. These significant gains are due to the following effects: D-D reactions, high burn-up efficiency of the resulting tritium and helium-3, significant transfer of nuclear spin from deuterium through D-D reactions to the resulting tritium and helium-3, and direct energy conversion efficiency that increases with spin polarization.

We have also argued that spin-polarized fuel could provide a path toward almost wholly aneutronic fusion in a D-${}^3$He plasma while still boosting the power density from D-${}^3$He fusion reactions. As long as D-D reactions are suppressed - resulting from the Quintet Suppression Factor having a value of zero - and the deuterium fuel remains fully polarized, D-${}^3$He reactions dominate. While we have not considered the effect of spin polarization on ${}^3$He-${}^3$He reactions, these reactions are aneutronic and the reactivity is orders of magnitude smaller than D-${}^3$He.

Spin-polarized fuel could also provide a path to tritium-free fusion power plants. This could be accomplished in two ways: (i) achieving an extremely high burn-up of tritium by increasing D-T fusion reactivity, (ii) suppressing the D+D$\to$p+T reaction through the Quintet Suppression Factor. Both of these pathways are highly speculative. 

Conversely, spin-polarized fuel could also provide a path to boosting the production of tritium and helium-3 in fusion power plants. A recent stellarator design proposed using beam-target D-D reactions for tritium production \cite{Swanson_2025}. Depending on the Quintet Suppression Factor and the subsequent spin transfer to tritium and helium-3, it may be possible to significantly boost tritium production and decrease tritium burn fraction in the plasma (estimated burn fraction of 0.09 of D-D tritons with unpolarized fuel in this design \cite{Swanson_2025}). The D-D tritium yield could be boosted by spin-polarized deuterium satisfying $P_\mathrm{D} = 1$, and the tritium burn-up rate could be decreased if tritium preferentially inherits spin satisfying $P_\mathrm{T} = -1/2$. In the most optimistic case where $P_\mathrm{D} = 1$, QSF = 2.5, and all of the tritium from D-D reactions satisfied $P_\mathrm{T} = -1/2$ so that the tritium burn fraction decreases to 0.045, recovered tritium would increase by a factor of 2.62 and the D-T neutron production rate would increase by a factor of 1.25.

Preserving the fuel polarization is one of the biggest uncertainties for spin-polarized fusion \cite{Kulsrud1986,Garcia2023,Baylor2023,Heidbrink2024,Garcia2025}. Curiously, the depolarization properties of spin-polarized fuel in D-${}^3$He plasmas may differ significantly to those in spin-polarized D-T plasmas. This difference could be particularly large when most of the fusion power in D-${}^3$He plasmas comes from D-D and its secondary reactions. In this case, maintaining the helium-3 polarization is relatively unimportant for the total fusion power, but the deuterium polarization is crucial. Because depolarization is a major concern, it may be easier to optimize the polarization lifetime of a single species (deuterium) than for multiple species (both deuterium and helium-3). Furthermore, even if preserving the polarization of both deuterium and helium-3 is important, the nuclear $g$-factor for helium-3 is negative \cite{Schneider_2022} whereas it is positive for deuterium \cite{Smaller_1951}. This is in contrast to a D-T plasma where the nuclear $g$-factors for both deuterium and tritium \cite{Neronov_2011} are positive. Because one of the most concerning depolarization mechanisms is resonance of the nuclear precession frequency ($\Omega_p = g \Omega_c$ where $\Omega_c$ is the cyclotron frequency) with circularly polarized plasma waves \cite{Kulsrud1986,Coppi_1983,Heidbrink2024}, the fact that the precession frequency has an opposite sign for deuterium and helium-3 could lead to different depolarization properties than for deuterium and tritium. Whether there is a significant difference remains to be investigated.

Despite these benefits, there are many outstanding questions to be addressed before D-${}^3$He would be preferable than D-T fusion. One of the biggest challenges is helium-3 supply \cite{Wittenberg_1992,Shea2010_helium,Simko2014_lunar,Meschini_2025}. While helium-3 is stable, it is very scarce on Earth. One possible solution to the helium-3 supply problem is to operate a D-${}^3$He plasma with a relatively low helium-3 fuel fraction -- as shown in this work, this can be desirable as sometimes low helium-3 fractions even maximize the fusion power. Such a plasma would be dominated by D-D reactions, whose tritium and helium-3 products would need a high burn efficiency to produce significant power. In low helium-3 fraction operation, the neutrons produced by D-D fusion may be sufficient to ensure helium-3 self-sufficiency by producing helium-3 via tritium production from lithium-6 or hydrogen neutron capture \cite{Ball_2019b}. However, this has the considerable drawback that the tritium half-life is 12 years, meaning a considerable tritium inventory would be needed to produce helium-3 for a helium-3-self-sufficient fusion power plant. Furthermore, if SPF were used to significantly boost the fusion power in D-${}^3$He plasmas -- largely by enhancing reactivity of the D-D reaction and increasing subsequent secondary burnup -- these benefits are also available to D-T plasmas. The main distinction between the D-${}^3$He and D-T plasmas would be 1) the challenges of handling helium-3 versus tritium 2) at fixed fusion power D-T plasmas would have more high-energy D-T neutrons, decreasing material lifetime and 3) D-${}^3$He plasmas might benefit from direct energy conversion.

Another challenge for generating energy from spin-polarized D-${}^3$He plasmas is the high uncertainty in the spin-polarized D-D reaction for (1) the cross section (through the Quintet Suppression Factor for aligned D-D spins), (2) the transfer of spin to the D-D fusion products, and (3) the velocity-space distribution of D-D products \cite{Lemaitre1993,Deltuva2007,Engels2003,Paetz2010}. Owing to a recent resurgence of interest in p-${}^{11}$B fusion \cite{Magee_2023} and the predicted 60\% enhancement of the p-${}^{11}$B fusion cross section with spin polarization \cite{Dmitriev_2006}, it may also be worthwhile to consider spin-polarized D-${}^3$He-${}^{11}$B plasmas, where the protons from D-${}^3$He fusion reactions could further fuel p-${}^{11}$B fusion reactions. Finally, we have used low-fidelity modeling in this work -- given the sensitivity of the fusion power to parameters such as secondary tritium and helium burn efficiency and the D-D quintet suppression factor, much higher fidelity is required validate the assumptions and results in this work.

\section{Acknowledgments} \label{sec:acknowledgements}

We are grateful for discussions with C. Galea. The US Government retains a non-exclusive, paid-up, irrevocable, world-wide license to publish or reproduce the published form of this manuscript, or allow others to do so, for US Government purposes. This work was supported by the U.S. Department of Energy under contract numbers DE-AC02-09CH11466, DE-SC0022270, DE-SC0022272.

\section{Data Availability Statement}

The data that support the findings of this study will be made openly available upon publication.

\appendix

\section{Fusion Reactions} \label{app:reactions}

\begin{figure*}[tb!]
    \centering
    \begin{subfigure}[t]{1.\textwidth}
    \centering
    \includegraphics[width=1.0\textwidth]{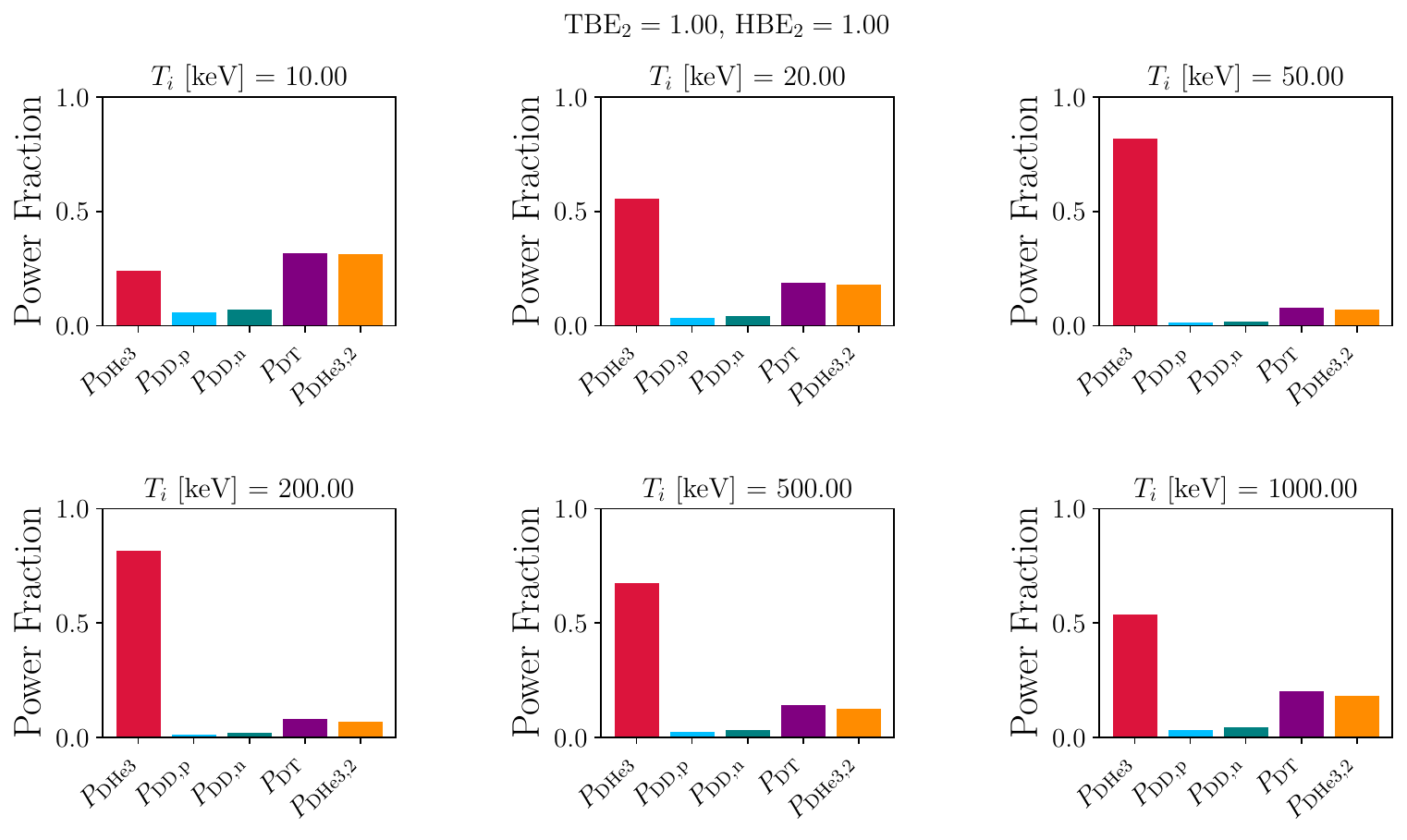}
    \end{subfigure}
    \centering
    \begin{subfigure}[t]{1.\textwidth}
    \centering
    \includegraphics[width=1.0\textwidth]{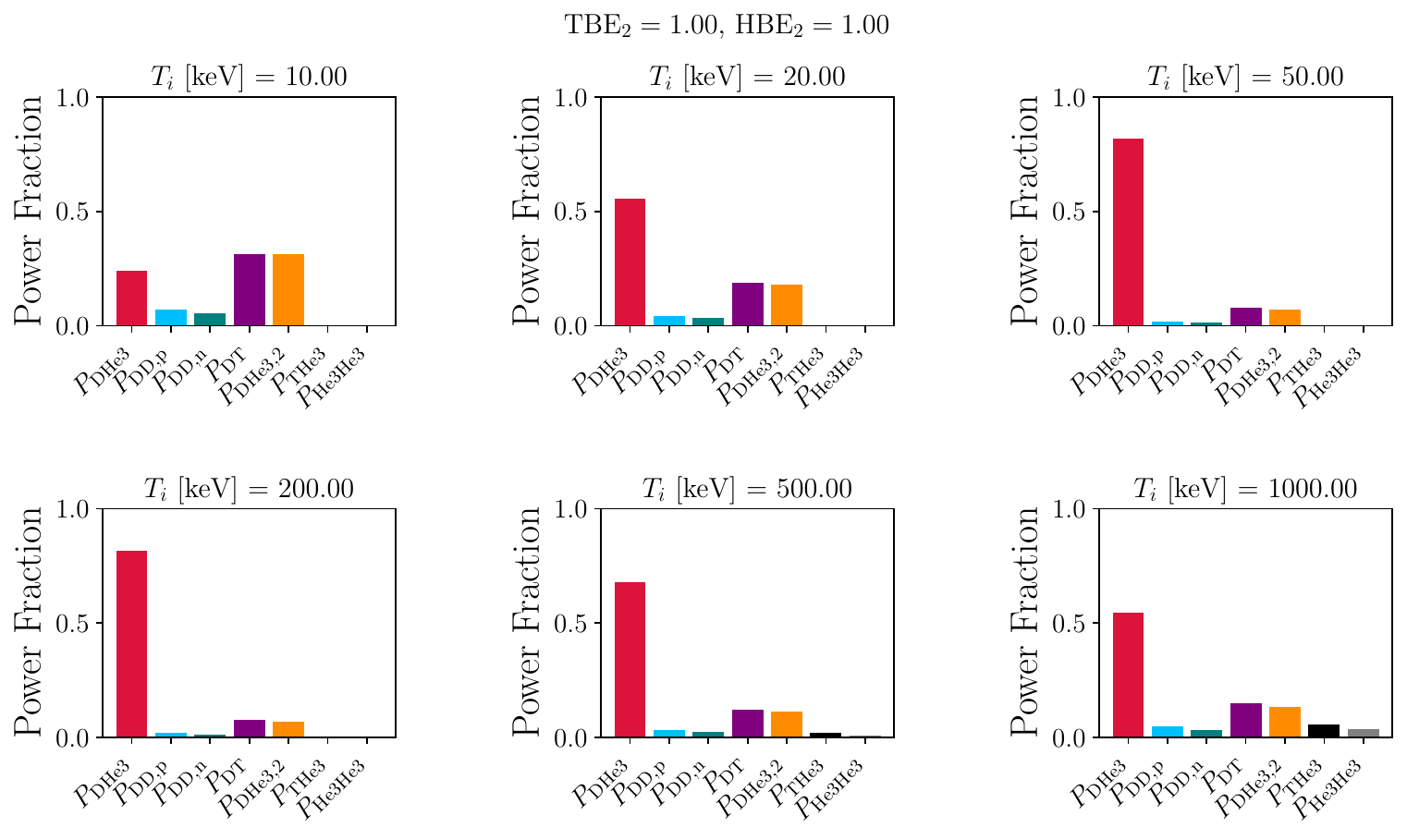}
    \end{subfigure} 
    \caption{Power fraction from different reactions. Top 6 plots corresponds to only including secondary D-T and D-${}^{3}$He reactions, whereas the bottom 6 plots also include T-${}^3$He and ${}^3$He-${}^3$He reactions. $n_\mathrm{{}^{3}He} = n_\mathrm{D}$.}
    \label{fig:extraspecies_comparison}
\end{figure*}

In this appendix we list the fusion reactions considered in this paper.
The deuterium-helium-3 fusion reaction
\begin{equation}
\mathrm{D} + \mathrm{{}^{3}He} \to \mathrm{H} \left( \mathrm{+14.7 \; MeV}  \right) + \mathrm{{}^4He} \left( \mathrm{+3.7 \; MeV}  \right)
\end{equation}
releases $E_\mathrm{D{}^{3}He} =$ 18.4 MeV of energy. The deuterium-tritium fusion reaction
\begin{equation}
\mathrm{D} + \mathrm{T} \to \mathrm{n} \left( \mathrm{+14.1 \; MeV}  \right) + \mathrm{{}^4He} \left( \mathrm{+3.5 \; MeV}  \right)
\end{equation}
releases $E_\mathrm{D{}^{3}He} =$ 17.6 MeV of energy. The deuterium-deuterium reaction has a neutronic and aneutronic branch. For unpolarized deuterium, both branches occur with 50\% probability. The neutronic deuterium-deuterium fusion reaction
\begin{equation}
\mathrm{D} + \mathrm{D} \to \mathrm{n} \left( \mathrm{+2.5 \; MeV}  \right) + \mathrm{{}^{3}He} \left( \mathrm{+0.8 \; MeV}  \right)
\end{equation}
releases $E_\mathrm{DD,n} =$ 3.3 MeV of energy. The aneutronic deuterium-deuterium fusion reaction
\begin{equation}
\mathrm{D} + \mathrm{D} \to \mathrm{H} \left( \mathrm{+3.0 \; MeV}  \right) + \mathrm{T} \left( \mathrm{+1.0 \; MeV}  \right)
\end{equation}
releases $E_\mathrm{DD,p} =$ 4.0 MeV of energy. The aneutronic ${}^3$He-${}^3$He fusion reaction
\begin{equation}
\mathrm{{}^{3}He} + \mathrm{{}^{3}He} \to 2 \mathrm{H} + \mathrm{{}^4He}
\end{equation}
releases $E_\mathrm{{}^{3}He{}^{3}He,p} =$ 12.9 MeV of energy. The tritium-helium-3 reaction has a neutronic and aneutronic branch. For unpolarized tritium and helium-3, the neutronic branch has a 59\% probability and the aneutronic branch a 41\% probability. The neutronic T-${}^3$He fusion reaction
\begin{equation}
\mathrm{T} + \mathrm{{}^{3}He} \to \mathrm{H} + \mathrm{n} + \mathrm{{}^4He}
\end{equation}
releases $E_\mathrm{THe3,p} =$ 12.1 MeV of energy. The aneutronic T-${}^3$He fusion reaction
\begin{equation}
\mathrm{T} + \mathrm{{}^{3}He} \to \mathrm{D} \left( \mathrm{+9.5 \; MeV}  \right) + \mathrm{{}^4He} \left( \mathrm{+4.8 \; MeV}  \right)
\end{equation}
releases $E_\mathrm{THe3,n} =$ 14.3 MeV of energy. In \Cref{fig:extraspecies_comparison} we show the effect of including T-${}^3$He and ${}^3$He-${}^3$He fusion reactions in a D-${}^3$He plasma. With the exception of very hot (T$\gtrsim$1000 keV) plasmas, the additional power from T-${}^3$He and ${}^3$He-${}^3$He fusion reactions is less than 1\% of the total fusion power.

\bibliographystyle{apsrev4-2} %
\bibliography{EverythingPlasmaBib} %

\end{document}